# Physical origin of giant excitonic and magneto-optical responses in two-dimensional ferromagnetic insulators

Meng Wu, Zhenglu Li, Ting Cao, and Steven G. Louie[*]

*Department of Physics, University of California at Berkeley, Berkeley, California 94720, USA*
*Materials Sciences Division, Lawrence Berkeley National Laboratory, Berkeley, California 94720, USA.*
[*]Email: sglouie@berkeley.edu

**Abstract**

The recent discovery of magnetism in atomically thin layers of van der Waals crystals has created great opportunities for exploring light-matter interactions and magneto-optical phenomena in the two-dimensional limit. Optical and magneto-optical experiments have provided insights into these topics, revealing strong magnetic circular dichroism and giant Kerr signals in atomically thin ferromagnetic insulators. However, the nature of the giant magneto-optical responses and their microscopic mechanism remain unclear. Here, by performing first-principles *GW* and Bethe-Salpeter equation calculations, we show that excitonic effects dominate the optical and magneto-optical responses in the prototypical two-dimensional ferromagnetic insulator, $CrI_3$. We simulate the Kerr and Faraday effects in realistic experimental setups, and based on which we predict the sensitive frequency- and substrate-dependence of magneto-optical responses. These findings provide physical understanding of the phenomena as well as potential design principles for engineering magneto-optical and optoelectronic devices using two-dimensional magnets.

**Introduction**

The magneto-optical (MO) effects, such as the magneto-optical Kerr effect (MOKE) and the Faraday effect (FE), have been intensively investigated experimentally in a variety of magnetic materials, serving as a highly sensitive probe for electronic and magnetic properties. Recent measurements using MOKE have led to the discovery of two-dimensional (2D) magnets, and demonstrated their rich magnetic behaviors [1, 2]. In particular, a giant Kerr response has been measured in ferromagnetic mono- and few-layer CrI$_3$ [2]. Magnetic circular dichroism (MCD) in photo absorption has also been measured in ferromagnetic monolayer CrI$_3$ [3]. However, the exact microscopic origin of such large MO signals and MCD responses in 2D materials is still unclear, because treating accurately sizable spin-orbit coupling (SOC) and excitonic effects that are essential for such an understanding in these systems has been beyond the capability of existing theoretical methods.

CrI$_3$, in its monolayer and few-layer form, is a prototypical 2D ferromagnetic insulator with an Ising-like magnetic behavior and a Curie temperature of about 45 K, exhibiting tremendous out-of-plane magnetic anisotropy [2]. Within one layer, the chromium atoms form a honeycomb structure, with each chromium atom surrounded by six iodine atoms arranged in an octahedron (Fig. 1a-b), and the point group of the structure is $S_6$. The crystal field therefore splits the Cr 3$d$ and I 5$p$ ligand states into $t_{2g}$ and $e_g$ manifolds; the spin degeneracy of which are further lifted by the exchange interaction. Although the major-spin $e_g$ states are delocalized due to strong $p$-$d$ hybridization, the magnetic moment is approximately 3$\mu_B$ at each Cr site, in accordance with an atomic picture from the first Hund's rule [4].

With our recently developed full-spinor *GW* and Bethe-Salpeter equation (BSE) methods, we show from first principles that the exceedingly large optical and MO responses in ferromagnetic monolayer CrI$_3$ arise *per se* from strongly bound exciton states consisting of spin-polarized electron-hole pairs that extend over several atoms. These exciton states are shown to have distinct characteristics compared with either the Frenkel excitons in ionic crystals and polymers, or Wannier excitons in other 2D semiconductors. By simulating realistic experimental setups, we further find that substrate configuration and excitation frequency of the photon strongly shape the MO signals. Our results provide the conceptual mechanism for the giant optical and MO responses, explaining quantitatively the recent experiments on CrI$_3$ [2,3]. In addition, comparison between bulk and monolayer CrI$_3$ reveals the pivotal role of quantum confinement in enhancing the MO signals.

**Results**

**Quasiparticle band structure.** An accurate first-principles calculation of the electronic structure of CrI$_3$ should account for both the dielectric polarization from the ligand groups and the on-site Coulomb interactions among the localized spin-polarized electrons. We adopt the following approach. The first-principles *GW* method has become a *de facto* state-of-the-art approach to describe dielectric screening and quasiparticles excitations in many real materials [5]. In practical calculations here, the $G_0W_0$ approximation [5] is employed where the *GW* self-energy is treated as a first-order correction, and the single-particle Green's function *G* as well as the screened Coulomb interaction *W* are calculated using eigenvalues and eigenfunctions from density-functional theory (DFT). Through the screened Coulomb interaction *W*, the nonlocal and dynamical screening effects as well as the self-energy effects beyond the DFT Kohn-Sham orbital energies (within the local-spin-density approximation (LSDA)) are captured. Also, in previous studies, the method of LSDA with an on-site Hubbard potential (LSDA+*U*) has served as a reasonable mean-field starting point for $G_0W_0$ calculations in correlated systems to avoid the spurious *p-d* hybridization [6,7]. In this work, we adopt an on-site Hubbard potential in the rotationally invariant formulation [8] with *U* = 1.5 eV and *J* = 0.5 eV, with fully-relativistic pseudopotentials and a plane wave basis set. The validity of this specific set of *U* and *J* has been carefully tested (see Supplementary Figure 1 and 2). Throughout this work, the magnetization of ferromagnetic monolayer CrI$_3$ is taken to be along the +*z* direction (Fig. 1b). As shown in Fig. 1c, our calculations reveal a strong self-energy correction to the quasiparticle bandgaps, due to the weak dielectric screening in reduced dimensions and the localized nature of the *d* states. The direct bandgap is 0.82 eV at the Γ point at the LSDA+*U* level, whereas the direct $G_0W_0$ quasiparticle bandgap including the self-energy effect is 2.59 eV, as shown in Fig. 1c. Throughout the calculations, we incorporate the SOC effect from the outset by employing full two-component spinor wave functions.

**Exciton-dominant optical responses.** The strong SOC strength and the ligand states strongly hybridizing with Cr *d* orbitals (see Supplementary Figure 3) have decisive influences on the electronic structure and optical responses of ferromagnetic monolayer CrI$_3$. SOC significantly modifies the bandgap and band dispersion near the valence band maximum [4]. Figure 2a shows the $G_0W_0$ band structure together with each state's degree of spin polarization (with an out-of-plane quantization axis), of which the orbital and spin degeneracy are consistent with the above discussions. After solving the first-principles BSE, which describes the electron-hole interaction [9], with spinor wave functions, we find a series of strongly bound dark (optically inactive) and bright (optically active) exciton states with excitation energies ($\Omega_S$) below the quasiparticle bandgap, as shown in the plot of the exciton energy levels (Fig. 2b). As seen in Fig. 2c, the calculated linearly polarized absorption spectrum including electron-hole interactions (i.e., with excitonic effects, solid red curve labeled $GW$-BSE) features three peaks at around 1.50 eV, 1.85 eV and 2.35 eV (below the quasiparticle gap of 2.59 eV), which are composed of several bright exciton states in each peak and denoted as A, B and C, respectively. This is in contrast to the calculated step-function-like noninteracting absorption spectrum (i.e., without excitonic effects, dashed blue curve labeled $GW$-RPA). The magnitude of the absorbance peak around 1.50 eV is deduced to be 0.7% from a previous differential reflectivity measurement (Fig. 2c, inset) [3], while our calculated absorbance with a broadening factor of 80 meV is around 0.6% at 1.50 eV. From our calculation (Fig. 2b), there are also two dark states (excitons D) with enormous binding energy of larger than 1.7 eV. The existence of two states of nearly the same energy comes from the fact that there are two Cr atoms in a unit cell. We plot the real-space exciton wave functions of these states, with the hole fixed on a Cr atom, in Fig. 2d-k. Unlike monolayer transition metal dichalcogenides where the bound excitons are of Wannier type with a diameter of several nanometers [10, 11], ferromagnetic monolayer $CrI_3$ hosts dark Frenkel-like excitons localized on a single Cr atom (Fig. 2d-e) and bright charge-transfer or Wannier excitons with wave functions extending over one to several primitive cells (Fig. 2f-k). These plots are consistent with the intuition that a smaller exciton binding energy is related to a larger exciton radius [11, 12]. Numerical calculations of the exciton radius further corroborate this conclusion (see Supplementary Table 1).

In addition, ferromagnetism and broken time-reversal symmetry (TRS) play vital roles in determining the internal structure of the exciton states in ferromagnetic monolayer $CrI_3$, in contrast to the Frenkel/charge-transfer excitons determined solely by flat-band transitions in boron nitride systems [13, 14], organic materials [15, 16] or alkali halides [9]. The eigenstate of an exciton is a coherent superposition of free electron-hole pairs at different $k$ points ($|cv, \mathbf{k}\rangle$), and may be written as $|S\rangle =$

$\sum_{c v \mathbf{k}} A_{c v \mathbf{k}}^{S} | c v, \mathbf{k} \rangle$, where $A_{c v \mathbf{k}}^{S}$ is the exciton envelope function in $k$-space [9]. Here $c$ denotes conduction (electron) states and $v$ denotes valence (hole) states. In Fig. 3a-d we plot the module square of the exciton envelope function in $k$-space. As expected of highly localized Frenkel excitons in real space, the lowest-lying dark state D in Fig. 3a shows a uniform envelope function in $k$-space, whereas the bright states A ($\Omega_S$ = 1.50 eV) and B$^+$ (at $\Omega_S$ = 1.82 eV) in Fig. 3b and 3c have the envelope function localized around Γ and have $s$ characters. From Fig. 3d, an interesting hexagonal petal pattern with a node at Γ can be found for exciton B$^-$ ($\Omega_S$ = 1.92 eV). In Fig. 3e-h, we plot the distribution of the constituent free electron-hole pairs specified by ($E_v$, $E_c$) for selected exciton states, weighted by the module squared exciton envelope function for each specific interband transition. It is obvious that the electron-hole composition of exciton D is distinct from those of the bright states (A and B).

Because of broken TRS and strong SOC effect [9], the electron (hole) states that compose a given exciton in this system are from Bloch wave functions with spin polarization along different directions, giving rise to rich excitonic spin configurations. In fact, the lowest-lying bound exciton states are all formed by Kohn-Sham orbitals with particular spin-polarization. Our calculations verify that the dark excitons D are dominated by (> 99.5%) transitions between the major-spin valence bands and minor-spin conduction bands. The bright states (forming peaks A, B and C) in Fig. 2f-k and Fig. 3b-d&f-h, however, are all dominated by (> 96%) transitions between the major-spin valence bands and major-spin conduction bands (see Supplementary Table 2). Ligand field theory can provide a qualitative understanding of the lowest-lying D and A exciton states of which the optical transitions mainly occur among the localized Cr $d$ orbitals [3, 17]. However, ligand field theory is insufficient to evaluate the oscillator strength of the excitons quantitatively. In addition, the coexistence of Frenkel and Wannier excitons in our system poses significant challenges to ligand field theory, while this excitonic physics can be fully captured by the first-principles $GW$-BSE method.

**MO effects from first principles.** The above-mentioned internal structures of the exciton states are essential for a deeper understanding of the MO responses. Note that all the irreducible representations of the double group $S_6^D$ are one-dimensional, which facilitates our analysis of optical selection rules for circularly polarized lights around the Γ point, as shown in Fig. 3i. For

1$s$-like bright states A and B$^+$ wherein the transition mainly happens between the topmost valence band and the major-spin $e_g$ manifold near the $\Gamma$ point, only one $\sigma^+$ circularly polarized transition is allowed among all the transitions, e.g., $\langle c_3|\hat{p}_+|v_1\rangle \neq 0$. Here $|v_1\rangle$ denotes the first valence state, $|c_3\rangle$ denotes the third conduction state and $\hat{p}_+ = \frac{-1}{\sqrt{2}}(\hat{p}_x + i\hat{p}_y)$ is the momentum operator in spherical basis with an angular momentum equal to 1; $\sigma^\pm$ denotes the circularly polarized light with the complex electric field amplitude along the direction of the spherical basis: $\mathbf{e}_\pm = \frac{\mp}{\sqrt{2}}(\mathbf{e}_x \pm i\mathbf{e}_y)$, where $\mathbf{e}_x$ ($\mathbf{e}_y$) is the unit vector along the $+x$ ($+y$) direction. This conclusion is further confirmed by our first-principles circularly polarized absorption shown in Fig. 3j. The 2$s$-like exciton B$^-$, unlike A and B$^+$, is dominated by $\sigma^-$ circularly polarized transitions. We quantify the MCD of absorbance by calculating the contrast, $\eta = \frac{\text{Abs}(\sigma^+) - \text{Abs}(\sigma^-)}{\text{Abs}(\sigma^+) + \text{Abs}(\sigma^-)}$, where $\text{Abs}(\sigma^\pm)$ denotes the absorbance of $\sigma^+$ and $\sigma^-$ circularly polarized light, respectively. $\eta$ is dominated by $\sigma^+$ circularly polarized light below 1.8 eV (Fig. 3k). If we flip the magnetization direction, $\eta$ will also flip sign at all frequencies, which agrees with the measured MCD of photoluminescence signals [3].

In the following, we investigate the MO Kerr and Faraday effects of ferromagnetic monolayer CrI$_3$. Previous studies have shown that both SOC and the exchange splitting should be present to ensure non-zero MO effects in ferromagnets [18-22], and recent calculations within an independent-particle picture using DFT have been carried out for the MO responses of monolayer CrI$_3$ [23]. The essence of a theoretical modeling of the MO effects lies in accurately accounting for the diagonal and off-diagonal frequency-dependent macroscopic dielectric functions, which are readily available from our $GW$-BSE calculations with electron-hole interaction included. We find that the above-discussed giant excitonic effects in ferromagnetic monolayer CrI$_3$ strongly modify its MO responses, leading to significantly different behaviors going beyond those from a treatment considering only transitions between noninteracting Kohn-Sham orbitals [23]. Here we shall only consider the most physically relevant measurement for 2D ferromagnets, namely, polar MOKE (P-MOKE) and polar FE (P-FE), where both the sample magnetization and the wave vectors of light are along the normal of the surface. In accordance to typical, realistic experimental setup, we consider a device of ferromagnetic monolayer CrI$_3$ on top of a SiO$_2$/Si substrate (the thickness of SiO$_2$ layer is set to 285 nm, and Si is treated as semi-infinitely thick) [2], as shown in Fig. 4a. For insulating SiO$_2$ with a large band gap (8.9 eV), we use its dielectric constant $\varepsilon_{\text{SiO}_2} = 3.9$ [24]. For silicon, we perform first-principles *GW* (at the $G_0W_0$ level) and *GW*-BSE calculations, and incorporate the frequency-dependence of the complex dielectric function $\varepsilon_{Si}(\omega)$ (see Supplementary Figure 4). Assuming an incident linearly polarized light, we calculate the Kerr (Faraday) signals by analyzing the polarization plane of the reflection (transmission) light, which is in general elliptically polarized with a rotation angle $\theta_K$ ($\theta_F$) and an ellipticity $\chi_K$ ($\chi_F$) (see Supplementary Figure 5). Here we adopt the sign convention that $\theta_K$ and $\theta_F$ are chosen to be positive if the rotation vector of the polarization plane is parallel to the magnetization vector, which is along the +*z* direction.

We find that the MO signals are very sensitive to the thickness of $SiO_2$ and to the photon frequency. As shown in Fig. 4c-d and Supplementary Figure 6, the thickness of $SiO_2$ layer will strongly affect the MO signals, due to the interference of reflection lights from multiple interfaces [2]. Such interference has been accounted for with our three-interface setup in Fig. 4a. To analyze the relation between MO signals and dielectric functions, we also consider a simpler two-interface setup. For a two-interface setup with semi-infinitely thick $SiO_2$ layer, the Kerr angle $\theta_K$ (Fig. 4d, solid blue curve), is related to Im[$\varepsilon_{xy}$] (Fig. 4b, dashed blue curve) and therefore resonant with the exciton excitation energies; the Kerr ellipticity $\chi_K$ (Fig. 4d, dashed red curve), on the other hand, is proportional to Re[$\varepsilon_{xy}$] (Fig. 4b, solid blue curve). For a two-interface model, $\theta_K$ is also found to be proportional to $n_0/(n_2^2 - n_0^2)$, where $n_0$ ($n_2$) is the refractive index for the upper (lower) semi-infinitely thick medium. Moreover, the $\theta_K$ and $\chi_K$ are connected through an approximate Kramers-Kronig relations, as expected from previous works [22, 25]. Because of this, close attention should be paid in interpreting MOKE experiments on 2D ferromagnets, where the substrate configuration significantly changes the behavior of the MOKE signals. The existing experimental data of $\theta_K$, however, only have a few excitation frequencies of photons available, e.g., 5±2 mrad at 1.96 eV for HeNe laser [2]. As shown in Fig. 4c, our simulations with a 285 nm $SiO_2$ layer in the three-interface setup achieve the same order of magnitude for $\theta_K$ around the MO resonance at ~1.85 eV, in good agreement with experiment. Based on the simulations, we also predict a sign change of $\theta_K$ around 1.5 eV. For photon energies higher than the quasiparticle band gap, the plasmon resonance along with a vanishing $\varepsilon_{xx}$ will nullify our assumptions of continuous waves [25, 26]. It is also possible to achieve an in-plane ferromagnetic structure with an external magnetic field [27, 28]. However, due to the broken $C_3$ symmetry therein, we expect the system to have diminished values of MO signals (in the same polar configurations) but to remain having excitons with large binding energies, as confirmed by our first-principles calculations (see Supplementary Figure 7).

**Effects of quantum confinement.** To further understand the effects of quantum confinement in 2D magnets, we compare the MO properties of ferromagnetic bulk and monolayer $CrI_3$. Interestingly, the calculated optical properties of bulk $CrI_3$ are also dominated by strongly bound excitons with optical absorption edge starting from 1.5 eV (in good agreement with experiment [3]), while the quasiparticle indirect bandgap is 1.89 eV and the direct bandgap at $\Gamma$ is 2.13 eV (see Supplementary Figure 8). Within a one-interface model of semi-infinitely thick bulk $CrI_3$, $\theta_K$ reaches a magnitude of 60 mrad at the resonances at around 1.7 eV and 2.0 eV (Fig. 4g), proportional to $\text{Re}[\varepsilon_{xy}]$ shown in Fig. 4f. To study the quantum confinement effect, we employ the P-FE setup shown in Fig. 4e, because P-FE in this setup is almost linear with respect to the ferromagnetic sample thickness and free from the substrate effects. Our calculated magnitude of the specific Faraday angle ($|\theta_F|$) of bulk $CrI_3$ is $(1.3 \pm 0.3) \times 10^3$ rad·cm$^{-1}$ at the excitation frequency of 1.28 eV, in agreement with the experimental value of $1.9 \times 10^3$ rad·cm$^{-1}$ at the same excitation frequency [29]. By extrapolating the bulk $\theta_F$ to the monolayer thickness [30], and comparing with that of suspended ferromagnetic monolayer $CrI_3$ as shown in Fig. 4h, we find that quantum confinement significantly enhances the MO response by a factor of 2.5 near 2.0 eV and introduces a redshift of 0.2 eV.

**Discussion**

In summary, from our first-principles calculations, we discover that the optical and MO properties of ferromagnetic monolayer $CrI_3$ are dominated by strongly bound excitons of charge-transfer or Wannier characters. A systematic modeling framework for P-MOKE and P-FE experiments is also developed, where we have shown that the MO signals exhibit a sensitive dependence on photon frequency and substrate configuration. These findings of the exciton physics in 2D magnets should shed light on design principles for future magneto-optical and optoelectronic devices, such as photo-spin-voltaic devices [31] and spin-injecting electroluminescence [32,33]. As a prototypical monolayer Ising magnetic insulator with a bandgap in an easily accessible optical range, ferromagnetic monolayer CrI$_3$ is also expected to be useful in high-speed and high-density flexible MO drives using van der Waals homostructures or heterostructures [27, 34].

**Methods**

**First-principles *GW* and *GW*-BSE calculations.** First-principles calculations of the electronic structure of ferromagnetic monolayer CrI$_3$ (as the mean-field starting point of the $G_0W_0$ and BSE studies) were performed at the DFT-LSDA level, as implemented in the Quantum ESPRESSO package [35], with parameters for the on-site Hubbard interaction $U$ = 1.5 eV and Hund's exchange interaction $J$ = 0.5 eV [8]. A slab model with a 16 Å vacuum thickness was adopted to avoid interactions between periodic images. We employed optimized norm-conserving Vanderbilt pseudopotentials including Cr 3$s$ and 3$p$ semicore states [36, 37]. The Kohn-Sham orbitals were constructed with plane-wave energy cutoff of 80 Ry. Experimental structure was used in the calculations for both the bulk and monolayer CrI$_3$, with the lattice constants: $a$ = 6.867 Å [30] (see Supplementary Table 3). SOC was fully incorporated in our calculations. The *GW* (at $G_0W_0$ level) and *GW*-BSE calculations, for the quasiparticle and optical properties, respectively, were performed using the BerkeleyGW package [38]. The dielectric cutoff was set to 40 Ry. We adopted a 6×6×1 grid with 6 subsampling points for calculating the dielectric function in ferromagnetic monolayer CrI$_3$ [39]. An 18×18×1 grid was then used for calculating the self-energy corrections. We treated the dynamical screening effect through the Hybertsen-Louie generalized plasmon-pole model [5], and the quasiparticle bandgap was converged to within 0.05 eV. The resulting quasiparticle band structure was interpolated with spinor Wannier functions, using the Wannier90 package [40]. Within our *GW*-BSE calculations, the exciton interaction kernel was interpolated from an 18×18×1 grid to a 30×30×1 grid using a linear interpolation scheme [9], and the transitions between 21 valence bands and 14 conduction bands were considered in order to converge the calculation of the transverse dielectric functions from the *GW*-BSE results. The *GW* (at $G_0W_0$ level) and *GW*-BSE calculations of ferromagnetic bulk CrI$_3$ used identical energy cutoffs and convergence thresholds as of monolayer CrI$_3$, and we adopted a 4×4×4 grid for calculating the dielectric function and self-energy corrections in bulk CrI$_3$. The *GW*-BSE calculations of bulk CrI$_3$ employed a coarse grid of 6×6×6 which was further interpolated to a fine grid of 10×10×10. In this work, we obtained the calculated dielectric function of a ferromagnetic monolayer in a supercell slab model by using a thickness of a monolayer CrI$_3$ of $d$ = $c_{bulk}$/3 = 6.6 Å (see

Supplementary Figure 9 and 10). Our calculations were performed for suspended CrI$_3$ in vacuum. Addition of an insulating substrate, such as fused silica or hexagonal boron nitride (hBN), introduces a small redshift of the exciton energies (estimated to be less than 0.1 eV for a hBN substrate, see Supplementary Figure 11), while the strong excitonic effects still dominate the optical and MO responses. Effects of the on-site Hubbard potential on single-particle energies were systematically investigated to reveal the strong *p-d* hybridization of the major-spin $e_g$ states (see Supplementary Figure 12 and Supplementary Table 4).

**Group theory analysis.** We analyzed the symmetry of wave functions in ferromagnetic monolayer CrI$_3$. Ferromagnetic monolayer CrI$_3$ has point group symmetry $S_6 = C_3 \otimes C_i$. The irreducible representations labeled in Fig. 3i are for the double group $S_6^D$ due to the presence of strong spin-orbit coupling. The notation of the irreducible representations follows previous works [41, 42].

**Data availability.** The data that support the findings of this study are available from the corresponding author upon reasonable request.

**Code availability.** The BerkeleyGW package is available from [berkeleygw.org](berkeleygw.org).

**Acknowledgements**

The work was supported by the Theory Program at the Lawrence Berkeley National Lab (LBNL) through the Office of Basic Energy Sciences, U.S. Department of Energy under Contract No. DE-AC02-05CH11231, which provided the *GW & GW*-BSE calculations and simulations and by the National Science Foundation under Grant No. DMR-1508412 and Grant No. EFMA-1542741, which provided for theoretical formulation and analysis of the MOKE simulations. Advanced codes were provided by the Center for Computational Study of Excited-State Phenomena in Energy Materials (C2SEPEM) at LBNL, which is funded by the U.S. Department of Energy, Office of Science, Basic Energy Sciences, Materials Sciences and Engineering Division under Contract No. DE-AC02-05CH11231, as part of the Computational Materials Sciences Program. Computational resources were provided by the DOE at Lawrence Berkeley National Laboratory's NERSC facility and the NSF through XSEDE resources at NICS.

**Author contributions**

S. G. L. conceived the research direction and M. W. proposed the project. M. W. developed the full-spinor methods/codes, carried out computations and wrote the manuscript; M. W., Z. L. and T. C. analyzed the data; S. G. L. directed the research, proposed analyses and interpreted results. All authors discussed the results and edited this manuscript.

**Competing financial interests**

The authors declare no competing interests.

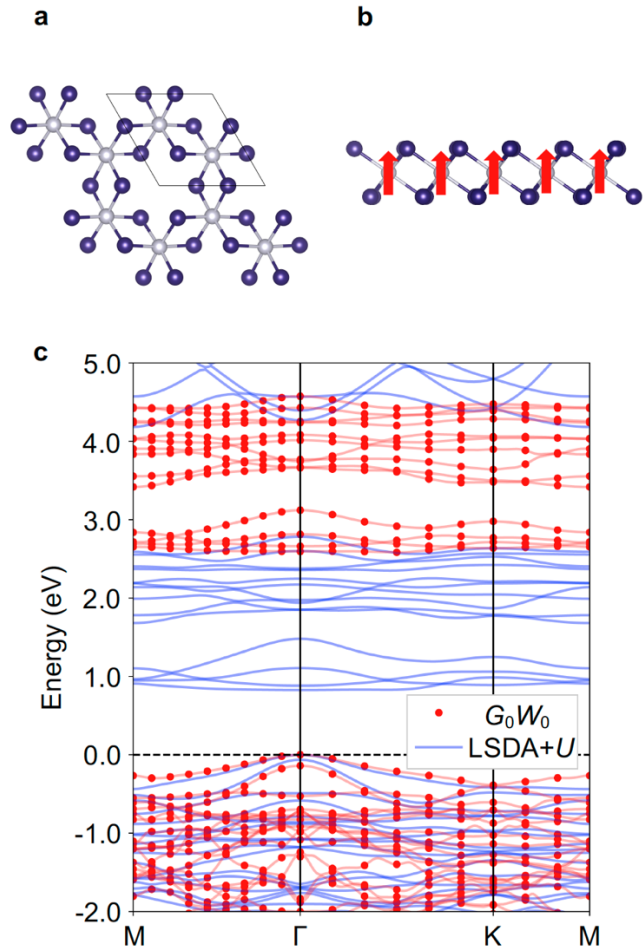

**Figure 1 | Crystal structure and electronic structure of ferromagnetic monolayer CrI$_3$. (a)** Crystal structure (top view) of monolayer CrI$_3$. Chromium atoms are in gray while iodine atoms in purple. **(b)** Crystal structure (side view) of ferromagnetic monolayer CrI$_3$. Red arrows denote the out-of-plane magnetization, which is pointing along the +z direction. **(c)** $G_0W_0$ (red dots) and LSDA+$U$ (blue lines) band structures of ferromagnetic monolayer CrI$_3$. A rotationally invariant Hubbard potential is employed with $U = 1.5$ eV and $J = 0.5$ eV in the LSDA+$U$ calculation, which is then used as the starting mean field for the $G_0W_0$ calculation. The $G_0W_0$ band structure is interpolated with spinor Wannier functions.

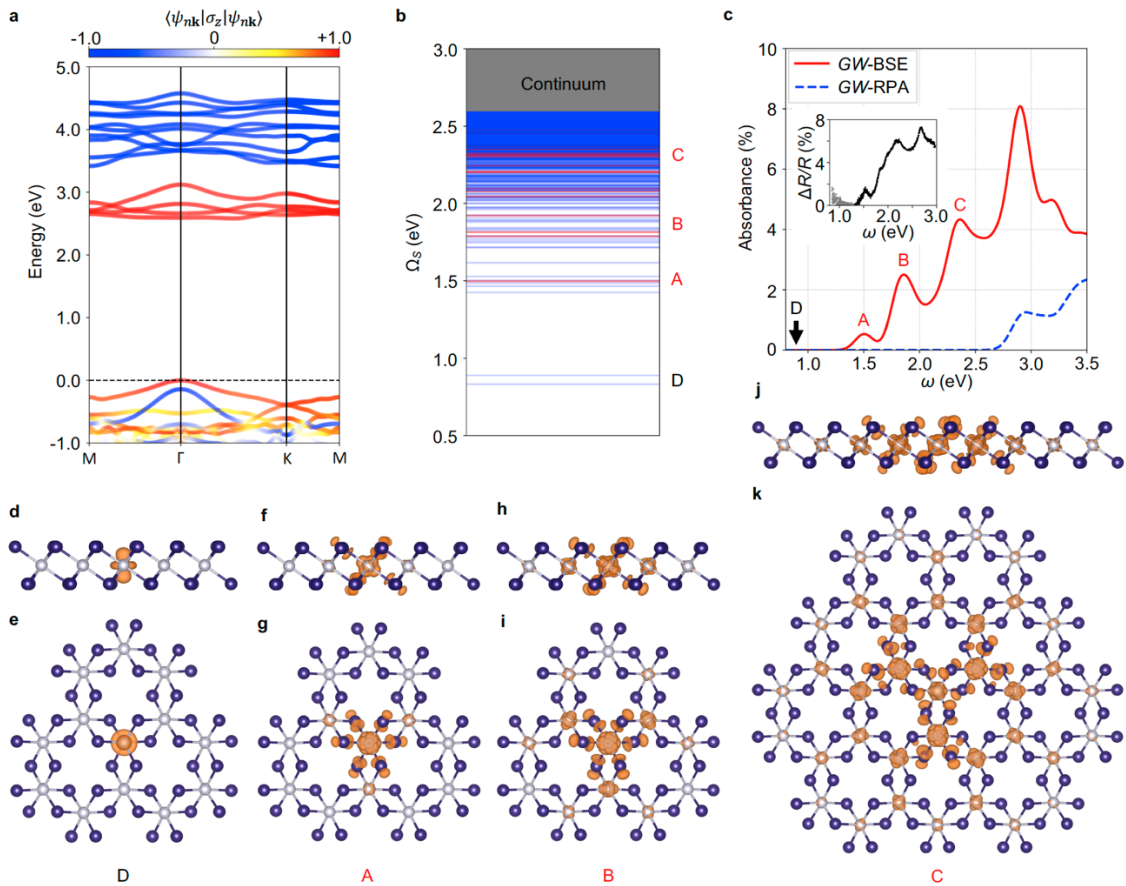

**Figure 2 | Calculated quasiparticle band structure and excitons in ferromagnetic monolayer CrI$_3$.** (**a**) $G_0W_0$ band structure with colors denoting the magnitude of spin polarization along the out-of-plane direction. The red (blue) color denotes the major-spin (minor-spin) polarization. (**b**) Exciton energy levels of monolayer CrI$_3$ calculated using the first-principles *GW*-BSE method. Optically bright exciton states are in red while dark ones in blue. The bright excitons have at least two orders of magnitude stronger oscillator strength compared with the dark ones. The free-electron-hole continuum starts from 2.59 eV. We label the bound exciton states with D for the lowest-lying dark states and A-C for the higher-lying bright states as evident in the plot of exciton levels. (**c**) Absorption spectrum of linearly polarized light with electron-hole interaction (*GW*-BSE, solid red line) and without electron-hole interaction (*GW*-RPA, dashed blue line). The inset data is extracted from ref. 3 showing the experimental differential reflectivity measured on a sapphire substrate, and the signals above 1.3 eV are shown in black for better comparison. (**d-k**) Exciton amplitudes in real space with the hole fixed on a Cr atom. Shown are iso-value surfaces of the amplitude square with the value set at 1% of the maximum value. Upper panel: side view. Lower

panel: top view. **(d-e)** Dark exciton D with an excitation energy $\Omega_S$ at 0.89 eV; **(f-g)** bright exciton A with $\Omega_S$ at 1.50 eV; **(h-i)** bright exciton B with $\Omega_S$ at 1.82 eV; **(j-k)** bright exciton C with $\Omega_S$ at 2.31 eV. Here the dominant states (with the largest oscillator strength among the nearby states in the same group) are plotted.

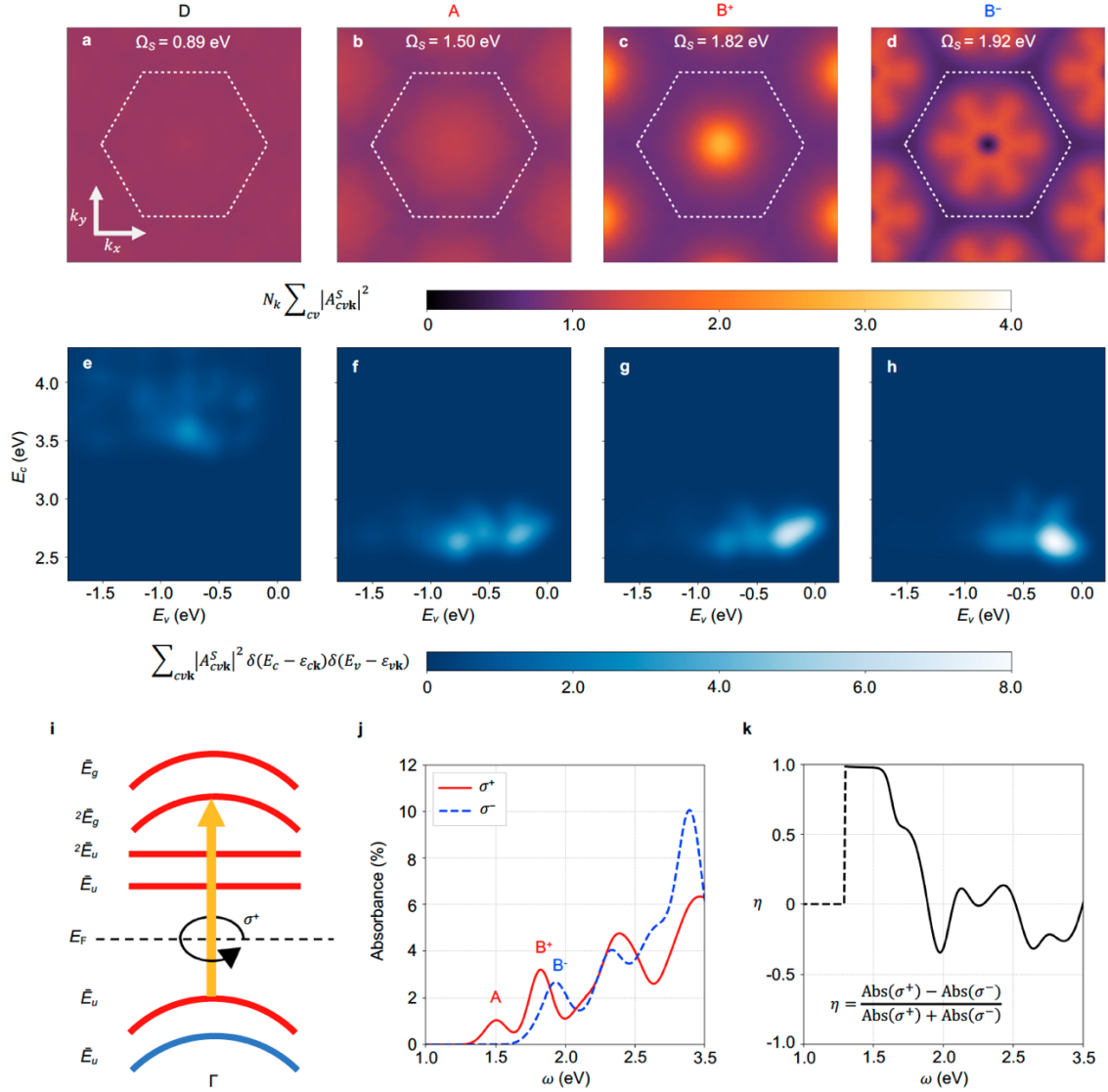

**Figure 3 | Internal structure of exciton states and MCD.** (**a-d**) Exciton envelope functions in *k*-space of (**a**) exciton D with $\Omega_S = 0.89$ eV, (**b**) exciton A with $\Omega_S = 1.50$ eV, (**c**) exciton B⁺ with $\Omega_S = 1.82$ eV and (**d**) exciton B⁻ with $\Omega_S = 1.92$ eV. The white dotted-line hexagon denotes the first Brillouin zone (BZ). The amplitudes are summed over band-pairs as given by $N_k \sum_{cv} |A^S_{cv\mathbf{k}}|^2$, where $A^S_{cv\mathbf{k}}$ describes the *k*-space exciton envelope function for the exciton state $|S\rangle$ and $N_k$ is the number of *k*-points in the first BZ. (**e-h**) The distribution of free electron-hole pair with electron energy at $E_c$ and hole energy at $E_v$ for selected exciton states: (**e**) exciton D, (**f**) exciton A, (**g**) exciton B⁺ and (**h**) exciton B⁻, weighted by module squared exciton envelope function for each interband transition between states $|v\mathbf{k}\rangle$ and $|c\mathbf{k}\rangle$, with quasiparticle energies $\varepsilon_{v\mathbf{k}}$ and $\varepsilon_{c\mathbf{k}}$, respectively. All the band energies are measured with respect to the valence band maximum. A

bivariate Gaussian energy broadening with equal standard deviation of 80 meV is used to smoothen the distribution. (**i**) Schematics of interband transitions around the $\Gamma$ point. The irreducible representations for Bloch states at the $\Gamma$ point are labeled. $E_F$ and the dashed line denote the Fermi level. The color scheme for spin polarization is the same as in Fig. 2a. Among all possible transitions in (**i**), only the indicated $\sigma^+$ circularly polarized dipole transition is allowed. (**j**) Frequency-dependent circularly polarized absorbance of ferromagnetic monolayer CrI$_3$ at normal incidence. The solid red (dashed blue) curve corresponds to the $\sigma^+$ ($\sigma^-$) circularly polarized light. (**k**) MCD of photo absorbance ($\eta$) as a function of the photon frequency. $\eta$ is set to zero below 1.3 eV as shown by the dashed line.

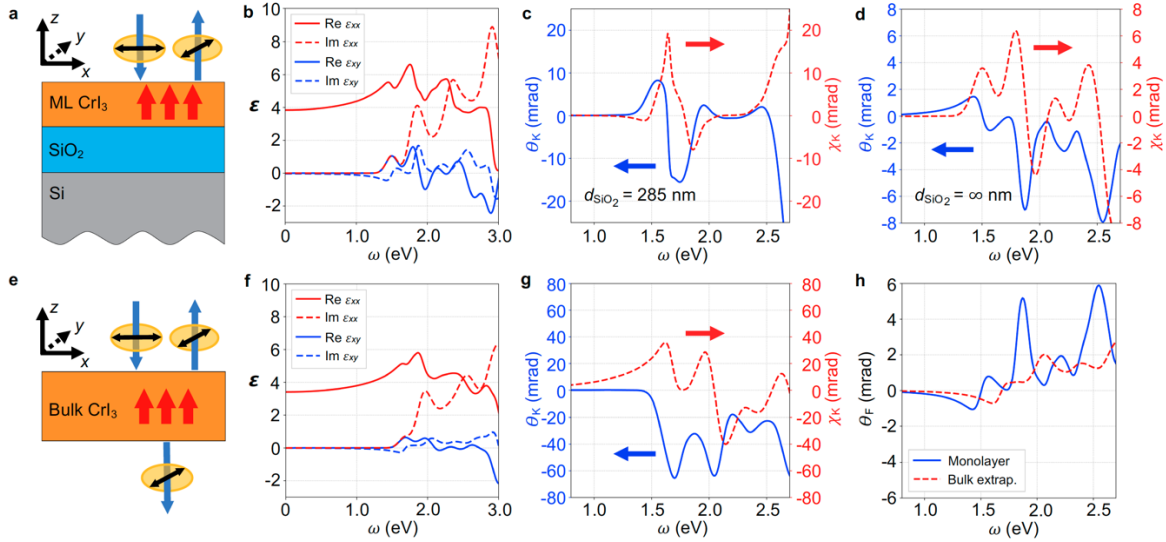

**Figure 4 | MO signals calculated from first-principles dielectric functions**. (**a**) P-MOKE setup consisting of layers of vacuum, ferromagnetic monolayer CrI$_3$, SiO$_2$ film, and semi-infinitely thick Si. Red arrows denote the out-of-plane magnetization, which is pointing along the +z direction. Blue arrows denote the propagation direction of light, and black double-headed arrows give the corresponding linear polarization direction. Each orange ellipse denotes a polarization plane of the electric field of light. (**b**) Calculated real part (solid lines) and imaginary part (dashed lines) of both the diagonal $\varepsilon_{xx}$ (red) and off-diagonal $\varepsilon_{xy}$ (blue) dielectric functions of ferromagnetic monolayer CrI$_3$, using a monolayer thickness $d = 6.6$ Å. An 80 meV energy broadening is applied. (**c**) Kerr angle $\theta_K$ (left, blue solid) and Kerr ellipticity $\chi_K$ (right, red dashed) for the P-MOKE setup with a 285 nm SiO$_2$ layer. (**d**) Kerr angle $\theta_K$ (left, blue solid) and Kerr ellipticity $\chi_K$ (right, red dashed) for the P-MOKE setup in (**a**) with semi-infinitely thick SiO$_2$ layer. (**e**) P-MOKE and P-FE setup of a suspended ferromagnetic bulk CrI$_3$ layer with the directions of light propagation and magnetization similar to (**a**). (**f**) Calculated real part (solid lines) and imaginary part (dashed lines) of both the diagonal $\varepsilon_{xx}$ (red) and off-diagonal $\varepsilon_{xy}$ (blue) dielectric functions of ferromagnetic bulk CrI$_3$, with an 80 meV energy broadening. (**g**) Kerr angle $\theta_K$ (left, blue solid) and Kerr ellipticity $\chi_K$ (right, red dashed) for the setup in (**e**) with infinitely thick ferromagnetic bulk CrI$_3$. (**h**) Comparison between Faraday angle $\theta_F$ of a suspended ferromagnetic monolayer CrI$_3$ and extrapolated bulk value down to the monolayer thickness (6.6 Å).

**Supplementary Information for "Physical origin of giant excitonic and magneto-optical responses in two-dimensional ferromagnetic insulators"**

Wu et al.

## A. $G_0W_0$ calculations using LSDA+$U$ as the starting point

In our $G_0W_0$ calculations, we will treat on the same footing the Hubbard potential ($V_{\text{Hub}}$) and the LSDA exchange-correlation potential $V_{xc}^{\text{LSDA}}$. That is, the self-energy correction is given by [1-4],

$$\Delta\Sigma = \Sigma - V_{xc} - V_{\text{Hub}}, \qquad (1)$$

where $\Sigma$ is the conventional self-energy operator in the $GW$ approximation. In this way, the Hubbard potential, together with the LSDA exchange-correlation potential, have been subtracted from the quasiparticle energies corrections to avoid the double-counting issue. As shown in Supplementary Figure 1 and 2, the $GW$ (at the $G_0W_0$ level) band structure and $GW$-BSE absorption spectrum exhibit only a weak dependence on $U$ and $J$ in a physically reasonable range of values [2-4].

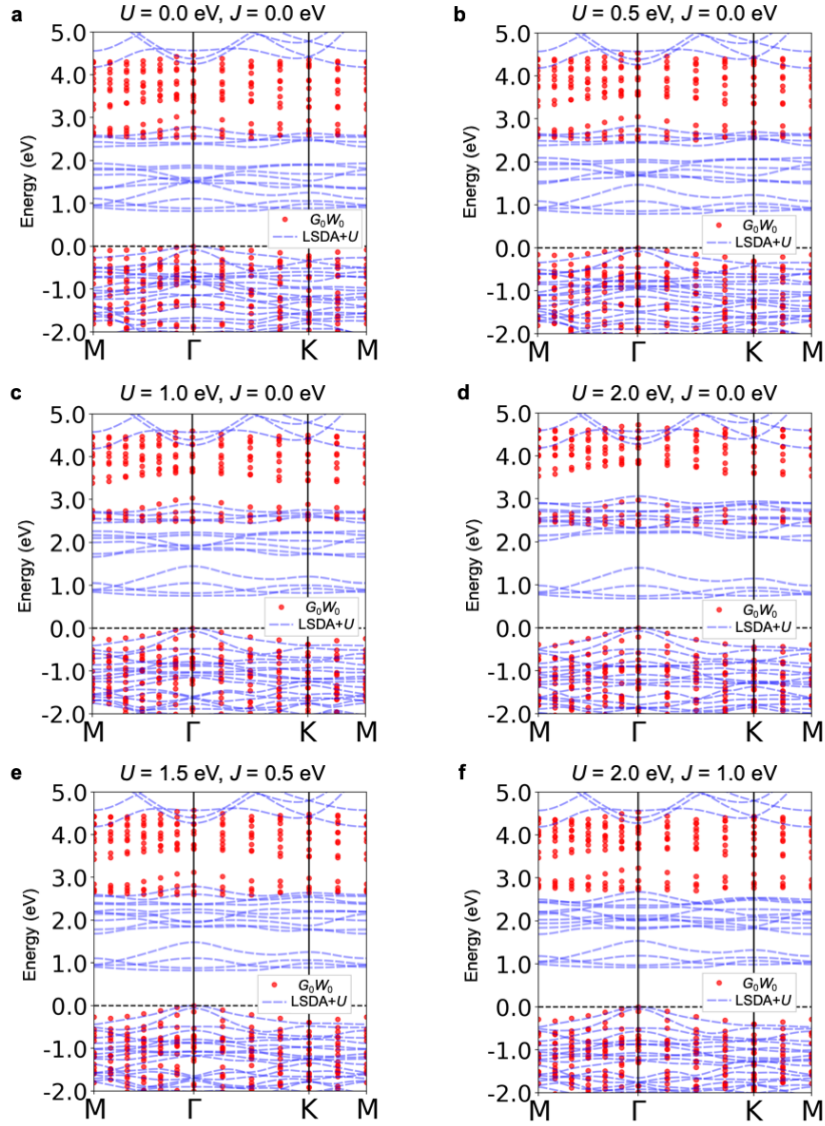

**Supplementary Figure 1** | Dependence of the $G_0W_0$ band structure on parameters of $U$ and $J$ within the LSDA+$U$ scheme. We consider the following sets of parameters: (**a**) $U = 0$ eV, $J = 0$ eV (**b**) $U = 0.5$ eV, $J = 0.0$ eV, (**c**) $U = 1.0$ eV, $J = 0.0$ eV, (**d**) $U = 2.0$ eV, $J = 0.0$ eV, (**e**) $U = 1.5$ eV, $J = 0.5$ eV, and (**f**) $U = 2.0$ eV, $J = 1.0$ eV.

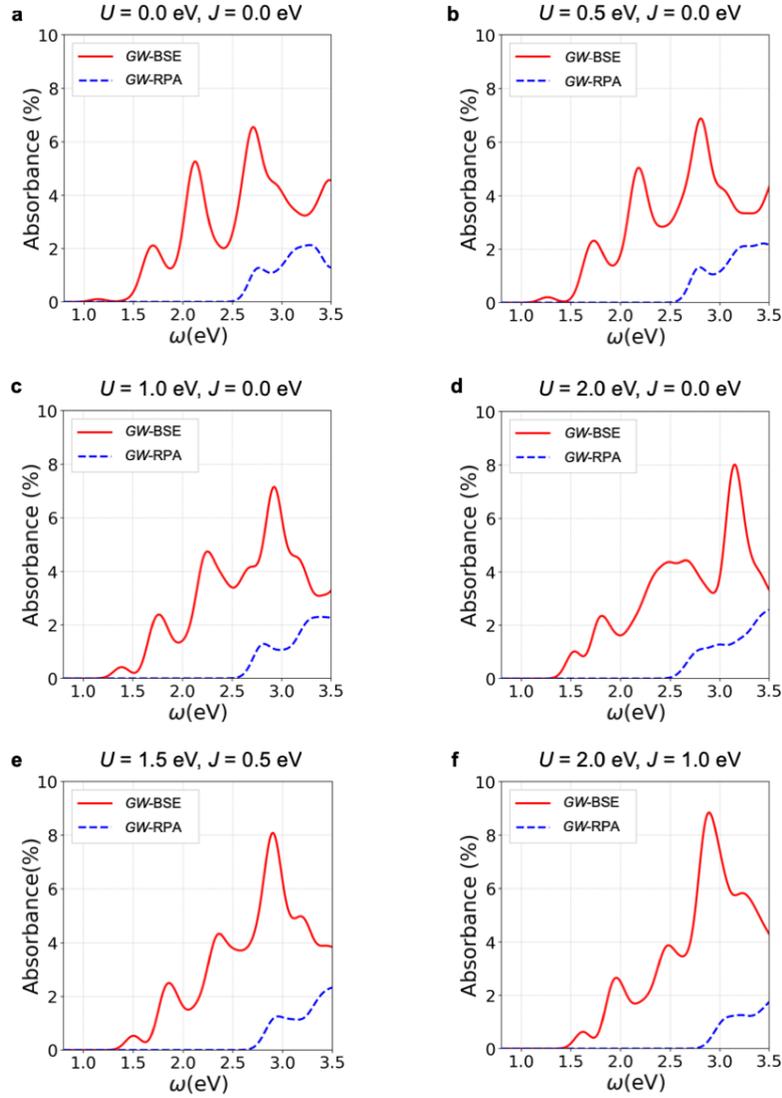

**Supplementary Figure 2 |** Dependence of the linear absorption spectrum at normal incidence on parameters of $U$ and $J$ within the LSDA+$U$ scheme, with ($GW$-BSE, solid red line) and without ($GW$-RPA, dashed blue line) electron-hole interaction. We consider the following sets of parameters: (**a**) $U = 0$ eV, $J = 0$ eV (**b**) $U = 0.5$ eV, $J = 0.0$ eV, (**c**) $U = 1.0$ eV, $J = 0.0$ eV, (**d**) $U = 2.0$ eV, $J = 0.0$ eV, (**e**) $U = 1.5$ eV, $J = 0.5$ eV, and (**f**) $U = 2.0$ eV, $J = 1.0$ eV.

## B. Density of states (DOS) of ferromagnetic monolayer $CrI_3$

The partial DOS (PDOS) based on the $G_0W_0$ band structure are shown in Supplementary Figure 3. We use spinor Wannier functions to interpolate the band structure as implemented in Wanner90 [5]. According to the calculated PDOS in Supplementary Figure 3, the first manifold of quasi-particle conduction bands (2.5 eV ~ 3.1 eV) consist of nearly equal contributions from Cr major-spin 3*d* and I major-spin 5*p* states, while the upper conduction bands (> 3.1 eV) are dominated by minor-spin Cr 3*d* states; and the top of valence bands host a decent amount of Cr major-spin 3*d* states, as well as the dominant I major-spin 5*p* states. In this way, the carrier energy distribution and the spatial localization of the dark state (Fig. 3a&e in the main text) origin from intra-atomic *d-d* transitions with spin flip. The contributions from occupied I 5*s* orbitals are around -12 eV below the highest occupied state, and those from occupied Cr 3s, 3*p* and 4*s* orbitals are below -40 eV.

It is obvious that, compared to LSDA, LSDA+*U* pushes the major-spin *d* states downwards in energy and significantly change the hybridization between the major-spin *d* states and the valence *p* states. However, LSDA+*U* has relatively small effects on the conduction states.

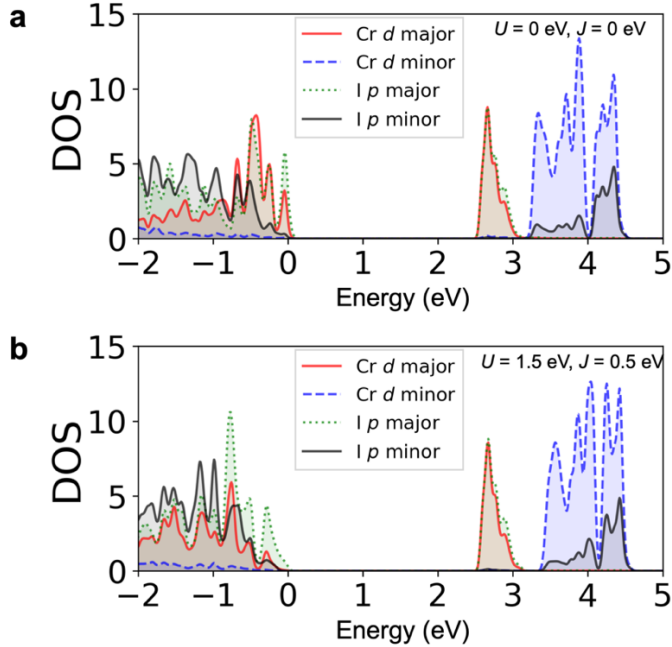

**Supplementary Figure 3** | Partial DOS of a ferromagnetic monolayer $CrI_3$ for (**a**) $U = 0$ eV, $J = 0$ eV and (**b**) $U = 1.5$ eV, $J = 0.5$ eV. The DOS is decomposed into contributions from Cr major-spin 3*d* orbitals (solid red curve), Cr minor-spin 3*d* orbitals (dashed blue curve), I major-spin 5*p* orbitals (green dotted curve) and I minor-spin 5*p* orbitals (black solid curve). The energy of the VBM is set to zero.

## C. Numerical evaluation of exciton radius

We use the exciton wave functions shown in Fig. 2 and Fig. 3 in the main text to evaluate the arithmetic mean radius $\langle |\mathbf{r}| \rangle$ and the root mean square radius $\sqrt{\langle r^2 \rangle}$ as shown in Supplementary Table 1. Our first-principles results agree with the intuition that a large binding energy indicates a small exciton radius, as inspired by a 2D hydrogenic model.

| Exciton states | A | B$^+$ | B$^-$ | C |
|---|---|---|---|---|
| $\Omega_S$ (eV) | 1.50 | 1.82 | 1.92 | 2.31 |
| $E_b$ (eV) | 1.09 | 0.77 | 0.67 | 0.28 |
| $\langle |\mathbf{r}| \rangle$ (Å) | 2.33 | 3.55 | 5.36 | 6.99 |
| $\sqrt{\langle r^2 \rangle}$ (Å) | 3.06 | 4.37 | 6.70 | 7.93 |

**Supplementary Table 1** | The arithmetic mean radius $\langle |\mathbf{r}| \rangle$ and root mean square radius $\sqrt{\langle r^2 \rangle}$ of selected bright exciton states. The excitation energy $\Omega_S$ and binding energy $E_b$ are also included for reference.

### D. Spin-decomposed exciton probability amplitudes

Since the two-component spinor wave functions are used in our calculations, it is possible for us to choose certain spinor components of wave functions and calculate the spin configuration of constituent carriers for each exciton state. To be specific, an exciton wave function in the spinor formalism can be written as,

$$\Psi(\mathbf{r}_e, \mathbf{r}_h) = \sum_{vc\mathbf{k}} A^S_{vc\mathbf{k}} \psi_{c\mathbf{k}}(\mathbf{r}_e) \psi^*_{v\mathbf{k}}(\mathbf{r}_h) := \sum_{vc\mathbf{k}} A^S_{vc\mathbf{k}} |c\rangle\langle v|$$

$$= \sum_{vc\mathbf{k}} A^S_{vc\mathbf{k}} \begin{pmatrix} |c\uparrow\rangle\langle v\uparrow| & |c\uparrow\rangle\langle v\downarrow| \\ |c\downarrow\rangle\langle v\uparrow| & |c\downarrow\rangle\langle v\downarrow| \end{pmatrix}. \quad (2)$$

The fractions for each spin configuration are listed in Supplementary Table 2.

| $\left|\sum_{vc\mathbf{k}} A^S_{vc\mathbf{k}}\|c\rangle\langle v\|\right|^2$ | $\|c\uparrow\rangle\langle v\uparrow\|$ | $\|c\uparrow\rangle\langle v\downarrow\|$ | $\|c\downarrow\rangle\langle v\uparrow\|$ | $\|c\downarrow\rangle\langle v\downarrow\|$ |
|---|---|---|---|---|
| $\Omega_S$=0.89 eV | 0.4% | 0.0% | 99.5% | 0.1% |
| $\Omega_S$=1.50 eV | 98.7% | 0.1% | 1.2% | 0.0% |
| $\Omega_S$=1.82 eV | 98.6% | 0.3% | 1.1% | 0.0% |
| $\Omega_S$=1.92 eV | 97.8% | 1.2% | 1.0% | 0.0% |
| $\Omega_S$=2.31 eV | 96.4% | 1.1% | 2.5% | 0.0% |

**Supplementary Table 2** | Decomposition of carrier spin configurations for selected exciton states.

**E. Computational details for Si and SiO$_2$**

The *GW* (at $G_0W_0$ level) and *GW*-BSE calculations of bulk Si are performed with the BerkeleyGW code [6]. The experimental lattice constant of $a$ = 5.43 Å at 300 K is adopted in the calculations. The resulting frequency-dependent dielectric function of Si is shown in Supplementary Figure 4. Note that $\varepsilon_{xy} = 0$ for bulk silicon and we could use a scalar dielectric function $\varepsilon = \varepsilon_{xx} = \varepsilon_{yy} = \varepsilon_{zz}$ for cubic crystals such as Si and SiO$_2$.

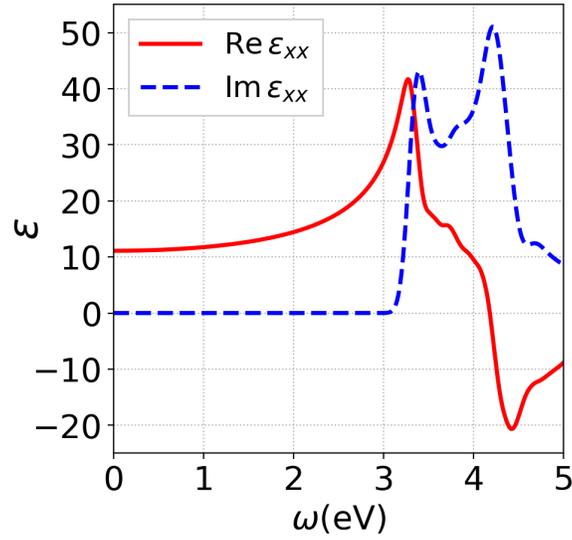

**Supplementary Figure 4** | Dielectric function $\varepsilon_{xx}$ of bulk Si from *GW*-BSE calculations. An 80 meV Gaussian broadening is employed.

The reflective index for bulk Si is calculated as,

$$n(\omega) = \sqrt{\varepsilon(\omega)}, \qquad (3)$$

where the static value $\mathrm{Re}[\varepsilon(\omega \to 0)] = 11.1$ is in excellent agreement with experimental value of 11.68 or 11.4 [7,8].

For bulk SiO$_2$, because the band gap (8.9 eV) is much larger than the energy range of interest in this problem (< 3.5 eV), we will only consider a static reflective index (with experimental value) with a very small imaginary part (~0.01$i$) to account for other dissipation channels [9],

$$n^2 = \varepsilon(\omega \to 0) = 3.9. \tag{4}$$

**F. Normal Modes**

In the P-MOKE configuration with at least $C_3$ rotational symmetry along the magnetization direction ($\mathbf{B} = B\hat{\mathbf{e}}_z$), the dielectric tensor takes the following form,

$$\varepsilon(\omega, \mathbf{B}) = \begin{pmatrix} \varepsilon_{xx}(\omega, \mathbf{B}) & \varepsilon_{xy}(\omega, \mathbf{B}) & 0 \\ -\varepsilon_{xy}(\omega, \mathbf{B}) & \varepsilon_{xx}(\omega, \mathbf{B}) & 0 \\ 0 & 0 & \varepsilon_{zz}(\omega, \mathbf{B}) \end{pmatrix}. \tag{5}$$

The Fresnel equation is given by,

$$[n^2 \mathbb{1} - \varepsilon - \mathbf{n}\!:\!\mathbf{n}] \cdot \mathbf{E} = 0, \tag{6}$$

where $\mathbf{n}$ is the complex refractory index,

$$\mathbf{n} = \frac{c\mathbf{k}}{\omega}. \tag{7}$$

After solving the Fresnel equations with the dielectric function in Supplementary Eq. 5 with $\mathbf{k} \parallel \hat{\mathbf{e}}_z$, we get the normal modes as the $\sigma^+$ and $\sigma^-$ circularly polarized plane waves, with distinct refractive indices,

$$n_\pm^2(\omega, B\hat{\mathbf{e}}_z) = \varepsilon_{xx}(\omega, B\hat{\mathbf{e}}_z) \pm i\varepsilon_{xy}(\omega, B\hat{\mathbf{e}}_z), \tag{8}$$

where the $+(-)$ in $n_\pm$ denotes the circularly polarized light with the complex electric field amplitude along the direction of the spherical basis:

$$\hat{\mathbf{e}}_\pm = \frac{\mp}{\sqrt{2}}(\hat{\mathbf{e}}_x \pm i\hat{\mathbf{e}}_y). \tag{9}$$

**G. Kerr signals and Faraday signals**

Following the convention in previous works [10], we get the ratio of complex amplitude of $\sigma^\pm$ circularly polarized reflection light through the ratio of the corresponding complex reflectivity $\tilde{r}^{(\pm)}$ as,

$$\frac{\tilde{E}^{(-)}{}_r}{\tilde{E}^{(+)}{}_r} = \frac{\tilde{r}^{(-)}}{\tilde{r}^{(+)}}. \tag{10}$$

Suppose that an incident linearly polarized light is along the $x$-axis, the relative complex amplitudes of the reflection light $\tilde{\mathbf{E}}_r$ (at fixed $z$ and $t$) along $x$- and $y$-axis are then given by,

$$\frac{\tilde{E}_{rx}}{\tilde{E}_{ry}} = \frac{1 + \tilde{r}^{(-)}/\tilde{r}^{(+)}}{i(1 - \tilde{r}^{(-)}/\tilde{r}^{(+)})}, \tag{11}$$

which defines an ellipse oriented slightly away from the $x$-axis as shown in Supplementary Figure 5b.

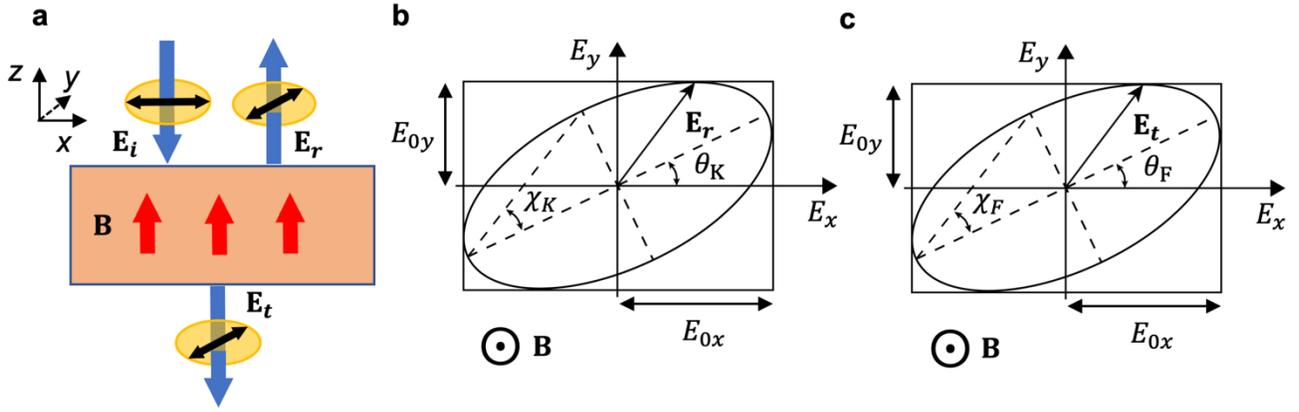

**Supplementary Figure 5** | (**a**) Configuration of the polar MO effects. The red arrows denote the magnetization of the sample, which is pointing along the +*z* direction. (**b**) The polarization plane of the reflection light. The polarization ellipse is oriented at a Kerr angle $\theta_K$ with respect to the *x*-axis. The Kerr ellipticity is defined through the ellipticity angle $\chi_K$ as shown. (**c**) The polarization plane of the transmission light. The polarization ellipse is oriented at a Faraday angle $\theta_F$ with respect to the *x*-axis. The ellipticity is defined through the Faraday ellipticity angle $\chi_F$.

We can calculate the Kerr angle $\theta_K$ and Kerr ellipticity $\chi_K$ as [11],

$$\tan 2\theta_K = \frac{2E_{0x}E_{0y}\cos\delta}{E_{0x}^2 - E_{0y}^2}, \quad -\frac{\pi}{2} < \theta_K \leq \frac{\pi}{2} \tag{12}$$

and

$$\sin 2\chi_K = \frac{2E_{0x}E_{0y}\sin\delta}{E_{0x}^2 + E_{0y}^2}, \quad -\frac{\pi}{4} < \chi_K \leq \frac{\pi}{4} \tag{13}$$

where $\delta = angle(\tilde{E}_{0y}/\tilde{E}_{0x})$, $E_{0x} = |\tilde{E}_{0x}|$ and $E_{0y} = |\tilde{E}_{0y}|$, and *angle(Z)* is a function that returns the phase angle of a complex number *Z*. A sign convention enters the expression of Kerr angle $\theta_K$: $\theta_K$ is chosen to be positive if the rotation vector of the polarization plane is parallel to the magnetization vector.

The Faraday angle $\theta_F$ and Faraday ellipticity $\chi_F$ are defined in a similar way for the transmission light, as shown in Supplementary Figure 5c.

**H. Multi-interface P-MOKE Setup**

Here we model the multi-interface P-MOKE setup in a systematic way [12, 13]. The goal is to calculate the complex reflection coefficients for $\sigma^\pm$ circularly polarized light, $\tilde{r}^{(\pm)} = \tilde{E}_r^{(\pm)}/\tilde{E}_i^{(\pm)}$ at the interface. The resulting complex reflection coefficients (at the topmost interface) for the three-interface model (shown in Fig. 4a in the main text) are given by,

$$\tilde{r}^{(\pm)} = \frac{e^{2i\delta_1^{(\pm)}}\left(\tilde{r}_{12}^{(\pm)} + e^{2i\delta_2}\tilde{r}_{23}\right) + e^{2i\delta_2}\tilde{r}_{01}^{(\pm)}\tilde{r}_{12}^{(\pm)}\tilde{r}_{23} + \tilde{r}_{01}^{(\pm)}}{1 + e^{2i(\delta_1^{(\pm)}+\delta_2)}\tilde{r}_{01}^{(\pm)}\tilde{r}_{23} + e^{2i\delta_1^{(\pm)}}\tilde{r}_{01}^{(\pm)}\tilde{r}_{12}^{(\pm)} + e^{2i\delta_2}\tilde{r}_{12}^{(\pm)}\tilde{r}_{23}}. \tag{14}$$

where the one-interface complex reflection coefficient between the *i*-th layer and the *j*-th layer is defined by $\tilde{r}_{ij} = (\tilde{n}_i - \tilde{n}_j)/(\tilde{n}_i + \tilde{n}_j)$, and the light path within the *i*-th layer is defined as $\delta_i = \omega\tilde{n}_i d_i/c$. The two-interface model can be achieved by taking $e^{i\delta_2} \to 0$. The one-interface model can be achieved by taking $e^{i\delta_1} \to 0$ and $e^{i\delta_2} \to 0$.

**I. Effects of substrate thickness on MO signals**

Here we investigate the influence of substrate thickness on MO signals. As described in the main text, we construct a three-interface P-MOKE setup with the order of vacuum-CrI$_3$-SiO$_2$-Si. We vary the thickness of the SiO$_2$ layer ($d_{SiO_2}$) and assume a semi-infinitely thick Si layer. We find that the interference of light reflected on each interface will be very sensitive to $d_{SiO_2}$. To be explicit, the amplitudes of the MOKE signals present a sensitive dependence on the substrate (Supplementary Figure 6), and it will be strongly modulated in the energy range of interest (1.0 ~ 3.5 eV) with increasing thickness of $d_{SiO_2}$, as shown in Supplementary Figure 6.

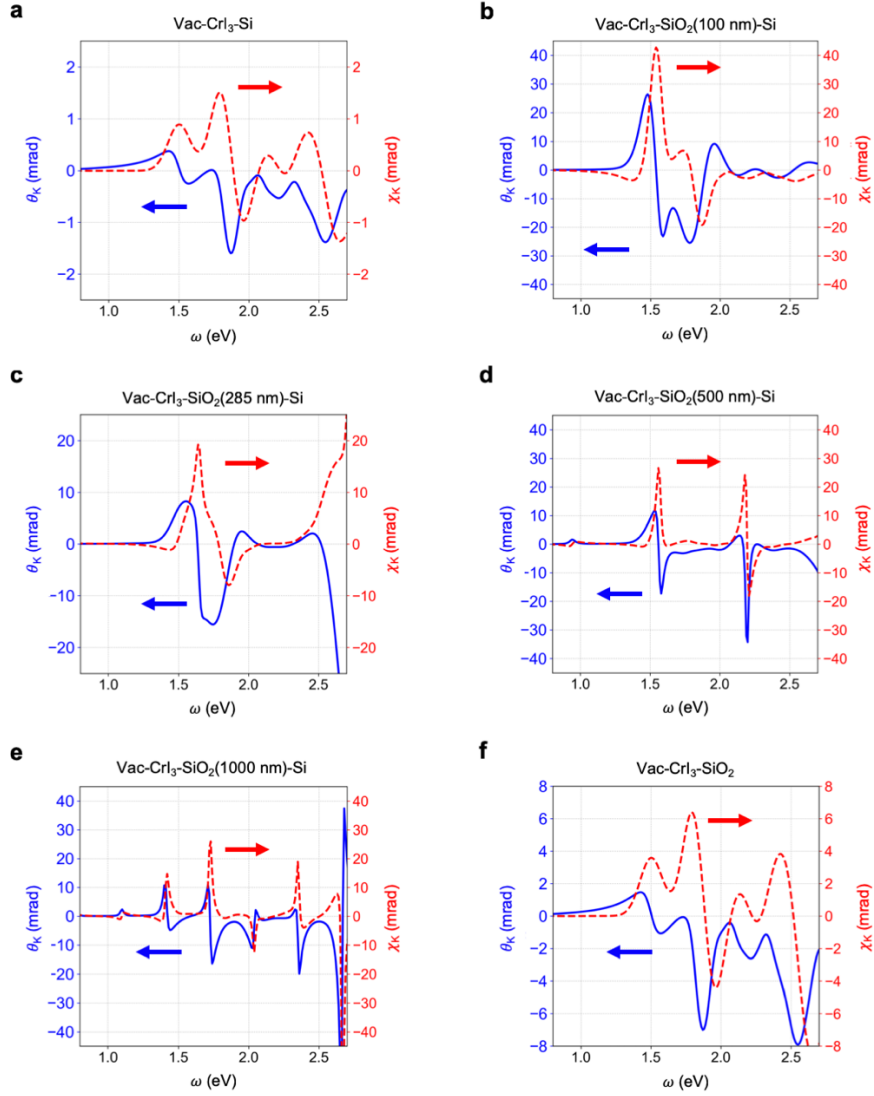

**Supplementary Figure 6** | Kerr angles $\theta_K$ (left, solid blue curve) and Kerr ellipticity $\chi_K$ (right, dashed red curve) for different P-MOKE setups with (**a**) vacuum-CrI$_3$-Si, (**b**) vacuum-CrI$_3$-SiO$_2$(100 nm)-Si, (**c**) vacuum-CrI$_3$-SiO$_2$(285 nm)-Si, (**d**) vacuum-CrI$_3$-SiO$_2$(500 nm)-Si, (**e**) vacuum-CrI$_3$-SiO$_2$(1000 nm)-Si and (**f**) vacuum-CrI$_3$-SiO$_2$ interfaces. All the CrI$_3$ layers in (**a-f**) refer to ferromagnetic monolayer CrI$_3$.

## J. In-plane ferromagnetic monolayer CrI$_3$

Our LSDA+$U$ and $G_0W_0$ calculations have reproduced the reported magneto band structure effect where relevant bands are degenerate at the $\Gamma$ point for the case of an in-plane ferromagnetic monolayer CrI$_3$ [14]. We have obtained an indirect bandgap of 2.64 eV at the $G_0W_0$ level in the in-plane polarized structure, which is only slightly smaller than the direct bandgap at the $\Gamma$ point (2.69 eV). The rotated magnetization has a strong impact on the polar MO signals, which can be understood from the symmetry. The dielectric function tensor $\boldsymbol{\varepsilon}$ is an axial tensor, which is in analogue to a dyad of two vectors. The broken $C_3$ rotational symmetry in an in-plane polarized structure leads to diminished amplitudes of $\varepsilon_{xy}$ and $\varepsilon_{yx}$, because the $x$- and $y$-components are no longer correlated. We therefore expect small polar MO signals in an in-plane polarized structure. Our first-principles $GW$-BSE calculations have confirmed the above analysis as shown in Supplementary Figure 7f regarding to the polar MO Faraday effect. Moreover, we find that there are still strong excitonic effects in the in-plane case. In fact, we could still identify the three excitonic peaks (A, B and C) below the quasi-particle bandgap (Supplementary Figure 7d), with the exciton A having a binding energy of 1.29 eV, even larger than that in the out-of-plane case. This is probably because the high band degeneracy in the in-plane case increases the joint density-of-states around the $\Gamma$ point, further enhancing the excitonic effects.

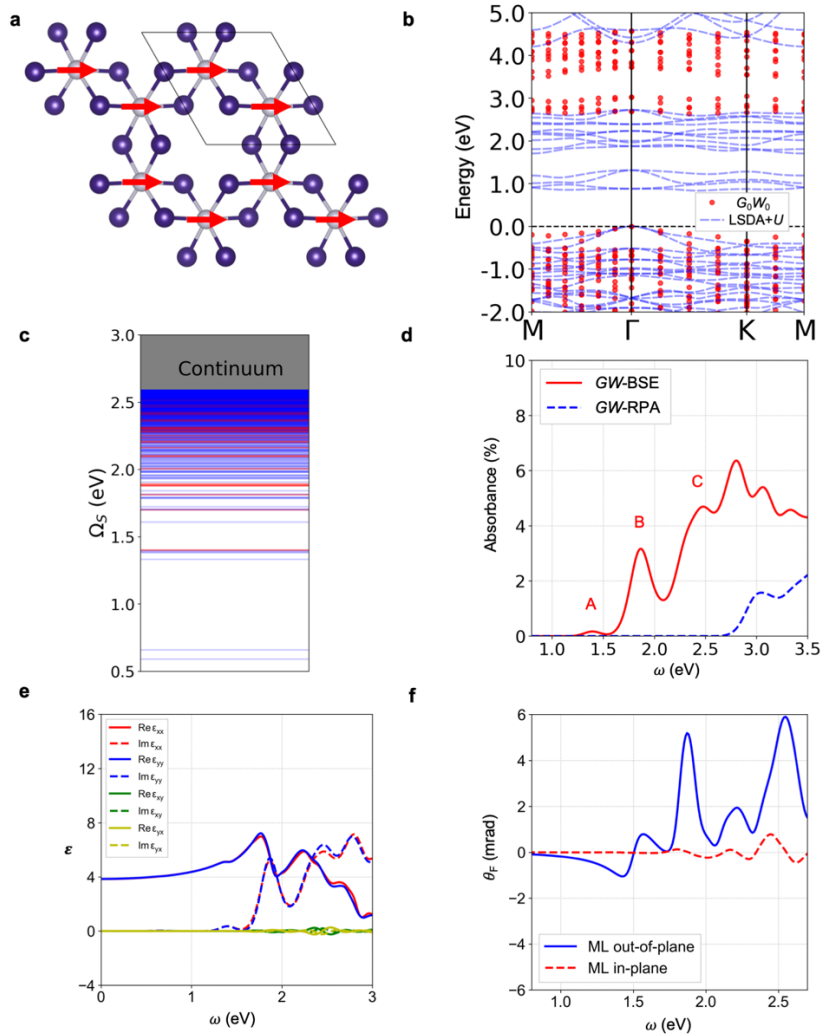

**Supplementary Figure 7 |** (**a**) Crystal structure and magnetization direction (red arrows) of an in-plane ferromagnetic monolayer CrI$_3$. (**b**) $G_0W_0$ (red solid) and LSDA+$U$ (blue dashed) band structure of the in-plane ferromagnetic monolayer CrI$_3$, where a Hubbard onsite potential with $U = 1.5$ eV & $J = 0.5$ eV is adopted. (**c**) Exciton levels of in-plane ferromagnetic monolayer CrI$_3$ from *GW*-BSE calculations. Bright excitons are colored in red while dark ones in blue. The continuum starts from 2.69 eV. (**d**) Linearly polarized absorption spectrum at normal incidence with (*GW*-BSE, solid red) and without (*GW*-RPA, dashed blue) electron-hole interactions. An 80 meV energy broadening is adopted in (**d**) and the following plots in (**e**) and (**f**). (**e**) Calculated real part (solid lines) and imaginary part (dashed lines) of $\varepsilon_{xx}$ (red), $\varepsilon_{yy}$ (green), $\varepsilon_{xy}$ (blue) and $\varepsilon_{yx}$ (yellow) dielectric functions of in-plane ferromagnetic monolayer CrI$_3$. (**f**) Comparison between Faraday angle $\theta_F$ of an out-of-plane ferromagnetic monolayer CrI$_3$ and an in-plane ferromagnetic monolayer CrI$_3$ with the same P-FE configuration as in Fig. 4e in the main text.

## K. Optical and MO properties of ferromagnetic bulk CrI$_3$

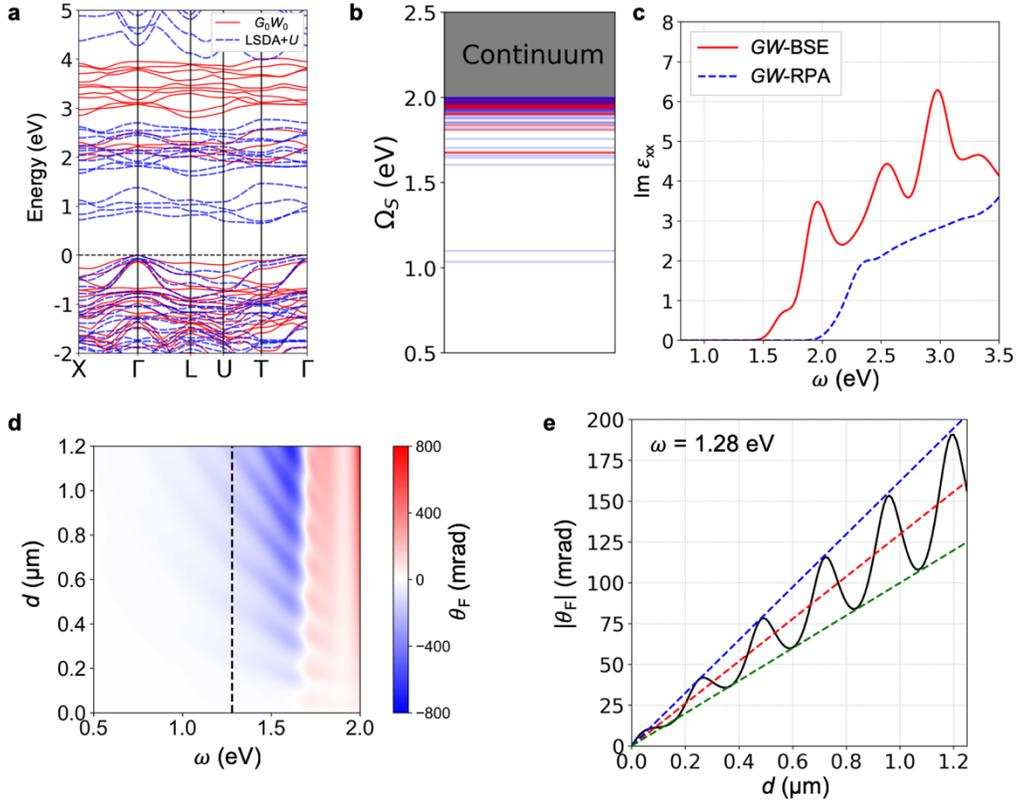

**Supplementary Figure 8** | (**a**) $G_0W_0$ (red solid) and LSDA+$U$ (blue dashed) band structure of ferromagnetic bulk CrI$_3$, where a Hubbard onsite potential with $U = 1.5$ eV and $J = 0.5$ eV is adopted. Each layer of bulk CrI$_3$ has the same out-of-plane direction as in monolayer CrI$_3$. The magnetization is along the out-of-plane direction. (**b**) Exciton levels of ferromagnetic bulk CrI$_3$ from $GW$-BSE calculations. Bright excitons are colored in red while dark ones in blue. The continuum starts from 2.0 eV. (**c**) Imaginary part of dielectric function $\varepsilon_{xx}$ of ferromagnetic bulk CrI$_3$ with (solid red) and without (dashed blue) electron-hole interactions. An 80 meV energy broadening is adopted in (**c**) and following plots. (**d**) Thickness and frequency dependence of Faraday angle $\theta_F$ for a vacuum-CrI$_3$-vacuum device, where $d$ refers to the thickness of the CrI$_3$ layer. (**e**) Thickness dependence of the amplitude of $\theta_F$ at $\omega = 1.28$ eV excitation, and the dashed lines are linear fits with slope: (blue) $1.6\times 10^3$, (red) $1.3\times 10^3$ and (green) $1.0\times 10^4$ mrad·cm$^{-1}$.

## L. Crystal structure and structure relaxation

We use the experimental structure for both ferromagnetic bulk and monolayer $CrI_3$. Bulk $CrI_3$ belongs to the space group $R\bar{3}$ (148) [15]. Both bulk and monolayer $CrI_3$ belong to the point group $S_6$. In fact, we have checked the validity of the employed pseudopotentials with first-principles structure relaxation. During the relaxation, we used a kinetic energy cutoff of 120 Ry. The van der Waals interaction is included in two ways: the rVV10 nonlocal density functional [16] and the semiempirical Grimme's DFT-D3 method [17], both of which have been implemented in the Quantum ESPRESSO package [18]. The structures have been fully relaxed until the force on each atom is less than 0.02 eV/Å. The spin-orbit coupling effects are not considered in the relaxation. We find that there is little deviation between the relaxed bulk structure and the experimental bulk structure. In addition, the lattice constants and internal coordinates barely change from bulk to monolayer structures, indicating much stronger intralayer bondings than interlayer bondings. For these reasons, we use the experimental structure for both bulk and monolayer calculations. The detailed structure parameters are listed in Supplementary Table 3.

|  | Parameters (Å) | $a$ | $c$ | Cr-I distance | Interlayer distance | Intralayer thickness |
|---|---|---|---|---|---|---|
| **Bulk** | Exp. [30] | 6.87 | 19.81 | 2.73 | 3.47 | 3.13 |
|  | rVV10 | 6.99 | 19.70 | 2.78 | 3.38 | 3.19 |
|  | DFT-D3 | 6.96 | 20.07 | 2.76 | 3.53 | 3.16 |
| **Monolayer** | rVV10 | 6.98 | N/A | 2.78 | N/A | 3.21 |
|  | DFT-D3 | 6.96 | N/A | 2.76 | N/A | 3.18 |

**Supplementary Table 3** | Structure parameters for ferromagnetic bulk and monolayer $CrI_3$. The lattice constants $a$ and $c$ refer to the conventional hexagonal cell.

The Quantum ESPRESSO structure input for bulk $CrI_3$ (with fractional coordinates) is listed below:

*CELL_PARAMETERS angstrom*
 *3.9648952386328364   0.000000000000000   6.602333333333333*
*-1.9824476193164182   3.433700000000000   6.602333333333333*
*-1.9824476193164182  -3.433700000000000   6.602333333333333*
*ATOMIC_POSITIONS crystal*
*Cr   0.666330330  0.666330330  0.666330330*

Cr  0.333669670  0.333669670  0.333669670
I  0.729082000  0.430059000  0.077768000
I  0.270918000  0.569941000  0.922232000
I  0.077768000  0.729082000  0.430059000
I  0.922232000  0.270918000  0.569941000
I  0.430059000  0.077768000  0.729082000
I  0.569941000  0.922232000  0.270918000

The Quantum ESPRESSO structure input for monolayer $CrI_3$ (with fractional coordinates) is listed below:

CELL_PARAMETERS angstrom
    6.8670000000    0.0000000000    0.0000000000
   -3.4335000000    5.9469964478    0.0000000000
    0.0000000000    0.0000000000   18.0000000000

ATOMIC_POSITIONS crystal
Cr  0.333333333  0.666666667  0.000377778
Cr  0.666666667  0.333333333  0.999622226
I  0.349896669  0.998803318  0.913106084
I  0.001196661  0.351093352  0.913106084
I  0.648906648  0.650103331  0.913106084
I  0.650103331  0.001196663  0.086893886
I  0.998803318  0.648906648  0.086893886
I  0.351093352  0.349896699  0.086893886

**M. Effect of a hexagonal boron nitride (hBN) substrate on exciton energies**

To study the effects of an insulating substrate, we perform *GW*-BSE calculations of a ferromagnetic monolayer $CrI_3$ on top of a monolayer hBN with an interlayer distance of 3.42 Å. The interlayer distance was determined with the van der Waals interaction through the DFT-D3 method [17]. It is expected that the substrate-induced exciton excitation energy redshift, albeit small, will quickly saturate with increasing thickness and additional layers of substrate will introduce negligible deviations [19-21]. Since hBN has a dielectric constant very similar to that of $SiO_2$ (both around 4), hBN and fused silica substrates of the same thickness should have similar screening effects on the exciton energies, as confirmed in previous work on transition metal dichalcogenides [19], as shown in Supplementary Figure 9.

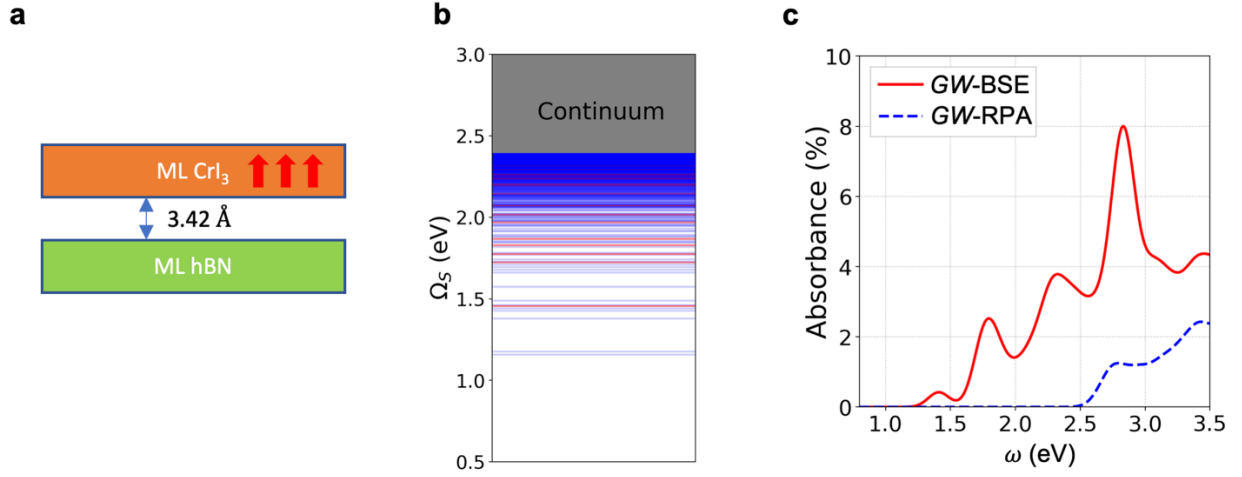

**Supplementary Figure 9** | (**a**) Schematic of a ferromagnetic monolayer CrI$_3$ on top of a monolayer hexagonal BN substrate, with an interlayer distance of 3.42 Å. A Hubbard onsite potential with $U$ = 1.5 eV & $J$ = 0.5 eV is adopted. (**b**) Exciton levels of ferromagnetic monolayer CrI$_3$ with a monolayer hBN substrate from $GW$-BSE calculations. Bright excitons are colored in red while dark ones in blue. The continuum starts from 2.41 eV. (**c**) Linearly polarized absorption spectrum at normal incidence with ($GW$-BSE, solid red) and without ($GW$-RPA, dashed blue) electron-hole interactions. An 80 meV energy broadening is adopted.

### N. Dielectric function for quasi-2D materials

As an extensive physical quantity, the dielectric function is ill-defined for quasi-2D materials. In this work, we rescale the calculated dielectric function in a slab model by the thickness of a monolayer CrI$_3$ ($d=c_{\text{bulk}}/3$=6.6 Å),

$$\varepsilon_{xx} = 1 + \frac{l}{d}(\tilde{\varepsilon}_{xx} - 1), \tag{15}$$

and

$$\varepsilon_{xy} = \frac{l}{d}\tilde{\varepsilon}_{xy}, \tag{16}$$

where $\tilde{\varepsilon}_{xx}$ and $\tilde{\varepsilon}_{xy}$ are calculated dielectric functions from the slab model (monolayer CrI$_3$ with vacuum) with thickness $l$ along the out-of-plane direction, and $\varepsilon_{xx}$ and $\varepsilon_{xy}$ are rescaled dielectric function for monolayer CrI$_3$ with monolayer thickness $d$.

To check the validity of our slab model calculations, we have performed calculations with different unit cell thickness $l$, and checked the scaling of the dielectric functions. As expected, we find a linear dependence of $(\varepsilon_{xx} - 1)$ and $\varepsilon_{xy}$ on $l$, which justifies our rescaling scheme. Also, it should be noted that, according to Eqn. (5-14) the calculated MO signals do not depend on the choice of $d$, because the final expression of MO signals depends only on $d\varepsilon_{xy}$, which is invariant with different choice of $d$. We have checked the validity of our slab model calculations in Supplementary Figure 10 and Supplementary Figure 11.

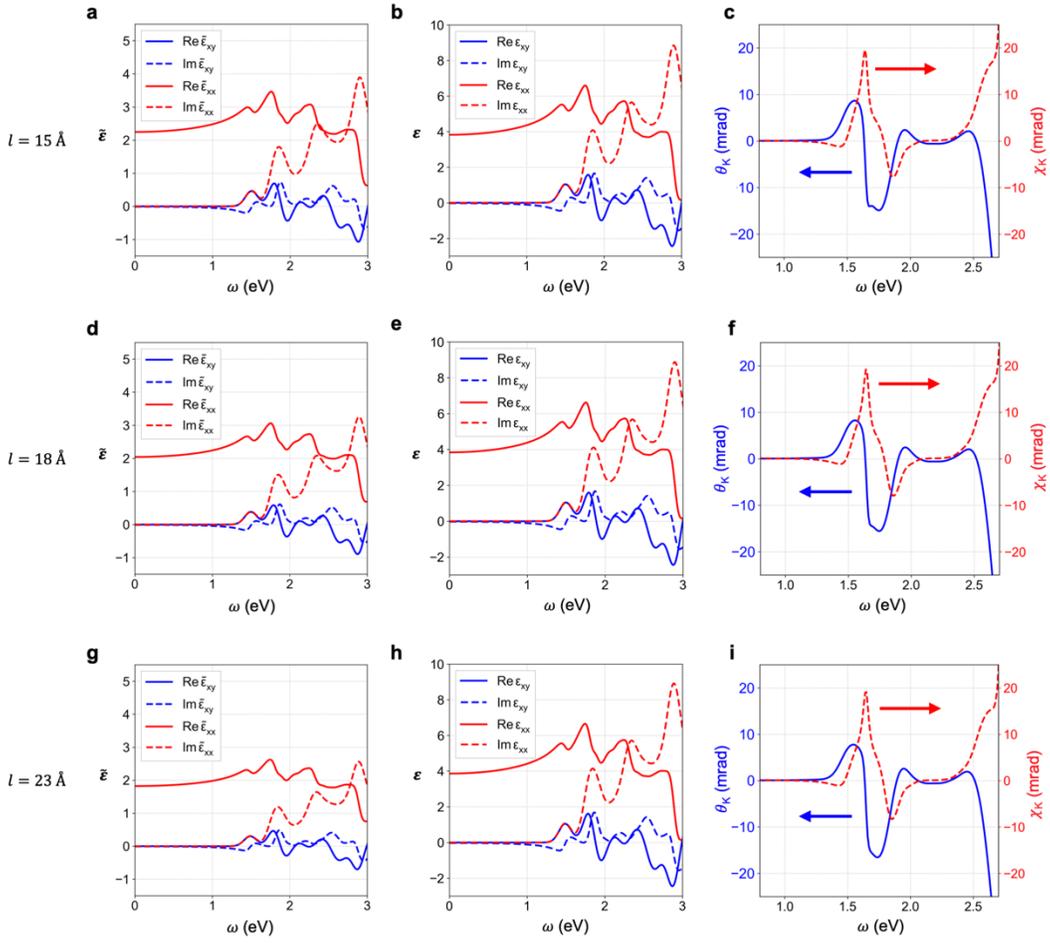

**Supplementary Figure 10** | Calculated dielectric functions $\tilde{\varepsilon}$ of ferromagnetic monolayer CrI$_3$ in a slab model at $GW$-BSE level for (**a**) $l = 15$ Å, (**d**) $l = 18$ Å and (**g**) $l = 23$ Å. Rescaled dielectric function $\varepsilon$ with $d = 6.6$ Å for (**b**) $l = 15$ Å, (**e**) $l = 18$ Å and (**h**) $l = 23$ Å. Kerr angles $\theta_K$ (left, solid blue curve) and Kerr ellipticity $\chi_K$ (right, dashed red curve) for the P-MOKE setup of vacuum-CrI$_3$-SiO$_2$(285 nm)-Si for (**c**) $l = 15$ Å, (**f**) $l = 18$ Å and (**i**) $l = 23$ Å. An 80 meV energy broadening is adopted.

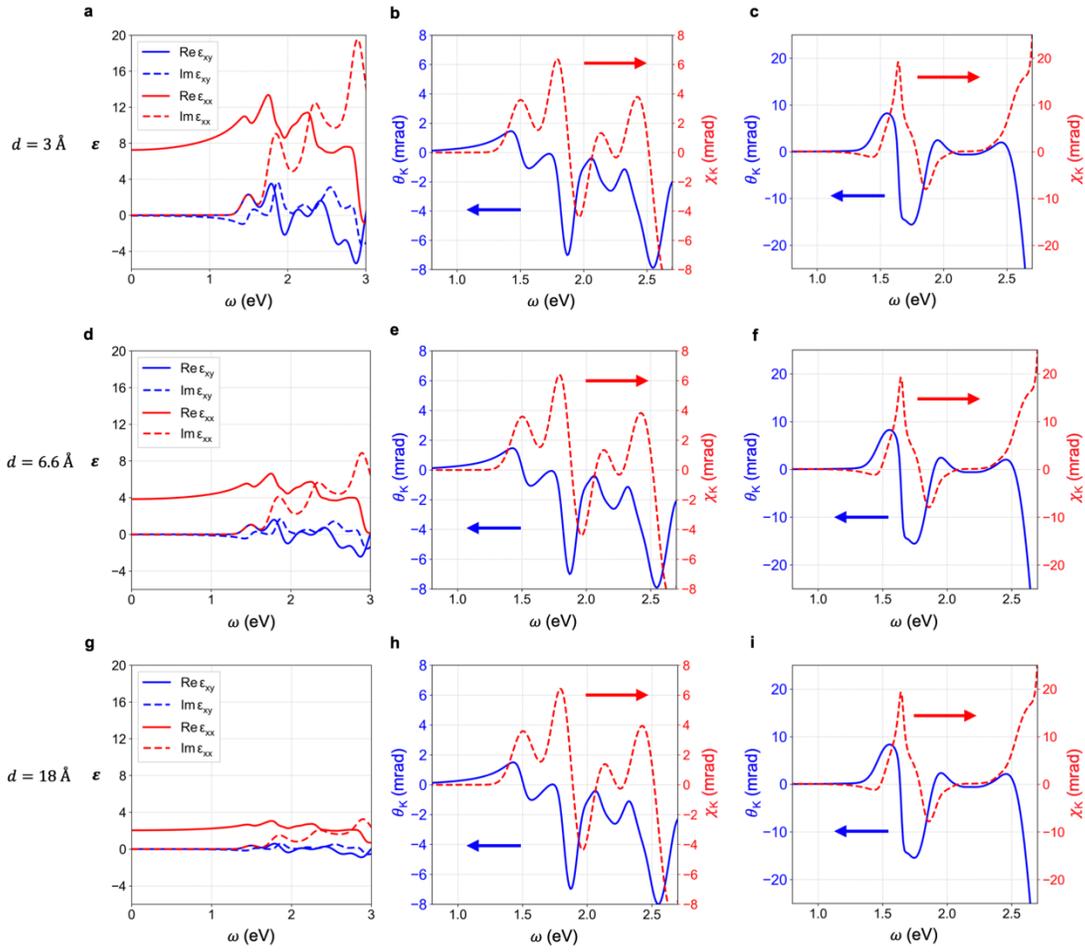

**Supplementary Figure 11** | Rescaled dielectric functions $\varepsilon$ of ferromagnetic monolayer CrI$_3$ in a slab model with $l = 18$ Å at $GW$-BSE level for (**a**) $d = 3$ Å, (**d**) $d = 6.6$ Å and (**g**) $d = 18$ Å. Kerr angles $\theta_K$ (left, solid blue curve) and Kerr ellipticity $\chi_K$ (right, dashed red curve) for the P-MOKE setup of vacuum-CrI$_3$-SiO$_2$ for (**b**) $d = 3$ Å, (**e**) $d = 6.6$ Å and (**h**) $d = 18$ Å. Kerr angles $\theta_K$ (left, solid blue curve) and Kerr ellipticity $\chi_K$ (right, dashed red curve) for the P-MOKE setup of vacuum-CrI$_3$-SiO$_2$(285 nm)-Si for (**c**) $d = 3$ Å, (**f**) $d = 6.6$ Å and (**i**) $d = 18$ Å. An 80 meV energy broadening is applied.

## O. Effect of $U$ on single-particle energies

There is an interesting behavior of LSDA+$U$ band energies with increasing $U$. As shown in the PDOS plots in Supplementary Figure 12, in contrast to the common impression of effects of +$U$, the bandgap and the first conduction peak (both of which contain substantial I $p$ orbital character) slightly decrease upon increasing $U$. Other Cr $d$ states, on the other hand, show expected behaviors by going away from the Fermi level with increasing $U$.

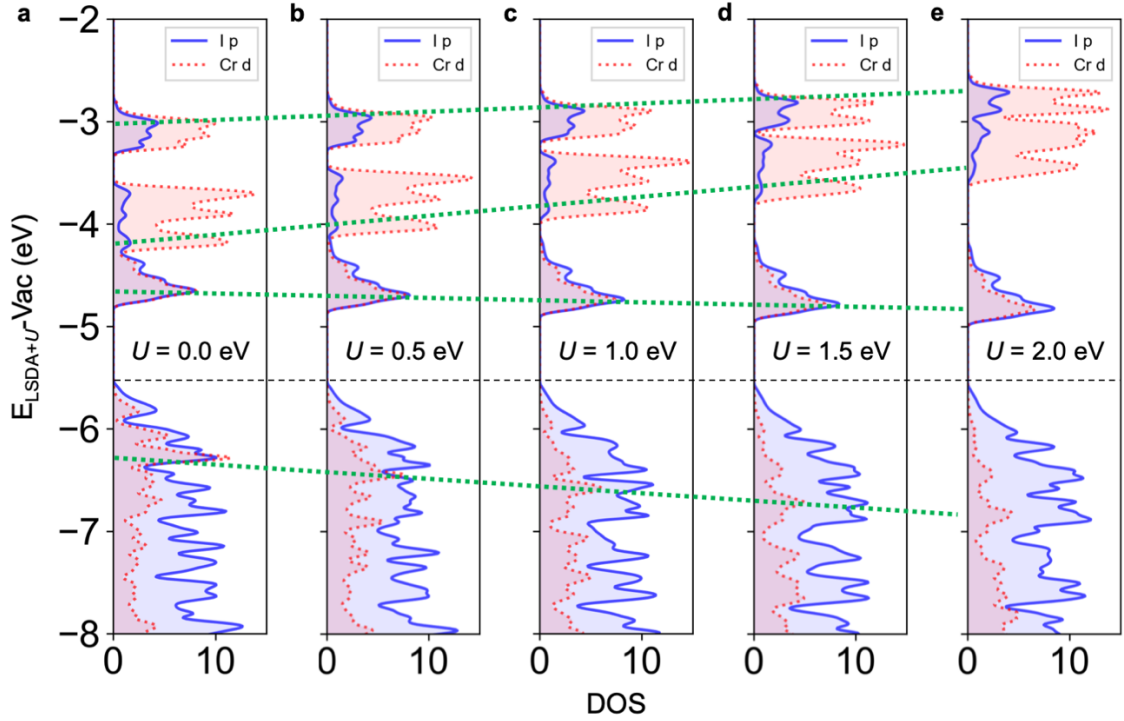

**Supplementary Figure 12** | PDOS plots of ferromagnetic monolayer CrI$_3$ at the LSDA+$U$ level with different $U$ values from (a) $U = 0$ eV to (e) $U = 2.0$ eV. $J$ is set to be zero in all cases for this illustrative investigation. The DOS is projected into contributions from Cr $d$ (red) and I $p$ (blue) orbitals. The green dots show the trends of band energies with increasing $U$. The single-particle energies have been aligned with the vacuum level in all the plots. The dashed black line indicates the VBM energy.

The effects of $U$ could be understood in terms of the *Janak theorem*: the eigenvalue is the derivative of the total energy with respect to the occupation of a state $i$, $\epsilon_i = \frac{\partial E_{\text{tot}}}{\partial n_i}$, where $i$ also includes the spin index [22]. In the case of LSDA+$U$, the total energy is given by $E_{\text{tot}} = E_{\text{LSDA}} + E_U - E_{\text{dc}}$ [1], and therefore the Bloch state eigenvalue can be expressed as,

$$\epsilon_{n\mathbf{k}} = \frac{\partial E_{\text{LSDA}}}{\partial n_{n\mathbf{k}}} + \frac{\partial (E_U - E_{\text{dc}})}{\partial n_{n\mathbf{k}}}. \tag{17}$$

To simplify our following discussions, we now assume $J = 0$. The energy correction from $+U$ can be estimated as,

$$\Delta\epsilon_{n\mathbf{k}}^{+U} = \sum_{at,m\sigma} U\left(\frac{1}{2} - n_{m\sigma}\right) |\langle m\sigma|\psi_{n\mathbf{k}}\rangle|^2, \tag{18}$$

where $n_{m\sigma}$ is the occupation number for the *diagonalized* and spin-polarized local basis $|m\sigma\rangle$, and the Bloch states $|\psi_{n\mathbf{k}}\rangle$ are assumed to be fixed before and after +$U$. To get $n_{m\sigma}$ and $|m\sigma\rangle$, we first use atomic wave functions (five spin-up $d$ orbitals and five spin-down $d$ orbitals for each Cr atom) as projectors to build the local orbital-resolved spin density $n_{ll'}^{\sigma}$ by summing over all the contributions from occupied Bloch states, with the quantization axis chosen along the $z$-axis. Here $l, l' = 1, ..., 5$ labels the $d$ orbitals and $\sigma = \uparrow, \downarrow$ labels the spin polarization. We then diagonalize this spin density and get the eigenvalues $n_m^{\sigma}$ and eigenvectors $|m\sigma\rangle$ as a linear combination of the five $d$ orbitals with spin polarization $\sigma$. In this way, the eigenvalue correction should incorporate an extra factor to quantify how much the projection of the Bloch state is onto the localized basis. This is in contrast to the atomic limit, $\Delta\epsilon_i^{+U} = U(\frac{1}{2} - n_i)$, which indicates that the occupied and unoccupied states at the LSDA level are further split by $U$. But there is strong $p$-$d$ hybridization in the CrI$_3$ systems, especially for $e_g$ states which form $\sigma$ bonds between Cr $d$ orbitals and I $p$ orbitals. In this way, the major-spin $e_g$ states become delocalized, which means the overlap between the first set of conduction Bloch states and the localized atomic orbital $|m\sigma\rangle$ used in LSDA+$U$ is small. To be explicit, $\Delta\epsilon_{n\mathbf{k}}^{+U}$ should be smaller for the $e_g$ states than for the $t_{2g}$ states, which has also been verified in our calculations. Moreover, the strong $p$-$d$ hybridization makes the local occupation $n_{m\sigma}$ deviate from the atomic limit: that is, 3 major-spin $d$ orbitals are completely occupied, while all the remaining $d$ orbitals are empty. Here we used $U = 2.0$ eV in a case study. As listed in Supplementary Table 4, the spin density has a more uniform distribution than that in the atomic limit.

| $U = 2$ eV & $J = 0$ eV | Minor spin | | | | | Major spin | | | | | $\Delta\epsilon_{+U}$ (eV) | |
|---|---|---|---|---|---|---|---|---|---|---|---|---|
| $n_{m\sigma}$ | 0.118 | 0.121 | 0.129 | 0.397 | 0.400 | 0.708 | 0.713 | 0.993 | 0.993 | 0.993 | Janak | LSDA+U |
| $|\langle n_{m\sigma}|v_1@\Gamma\rangle|^2$ | 1.1E-5 | 2.4E-5 | 8.3E-5 | 1.1E-5 | 1.3E-4 | 1.6E-9 | 3.0E-5 | 6.1E-3 | 1.6E-5 | 1.2E-2 | -0.04 | -0.01 |
| $|\langle n_{m\sigma}|c_1@\Gamma\rangle|^2$ | 7.4E-5 | 2.0E-6 | 6.6E-4 | 1.5E-6 | 1.2E-4 | 3.1E-7 | 2.2E-4 | 6.7E-2 | 1.6E-4 | 8.6E-2 | -0.30 | -0.16 |
| $|\langle n_{m\sigma}|c_5@\Gamma\rangle|^2$ | 1.6E-1 | 1.6E-1 | 5.4E-2 | 6.5E-3 | 2.4E-2 | 6.4E-4 | 1.0E-5 | 1.0E-5 | 1.2E-6 | 1.2E-6 | 0.58 | 0.63 |
| $|\langle n_{m\sigma}|c_{14}@\Gamma\rangle|^2$ | 1.9E-2 | 7.0E-3 | 3.6E-2 | 3.2E-6 | 2.0E-1 | 5.9E-6 | 2.4E-6 | 2.1E-5 | 9.1E-7 | 2.1E-5 | 0.17 | 0.27 |

**Supplementary Table 4 |** Occupation of diagonal basis on each Cr atom of a ferromagnetic monolayer CrI$_3$ from LSDA+$U$ calculations with $U = 2.0$ eV & $J = 0$ eV. The local orbital-resolved spin density matrix $n_{ll'}^{\sigma}$ is diagonalized with eigenvalues $n_{m\sigma}$ as shown in the second row. The projections of the 1$^{st}$

valence state and the 1$^{st}$, 5$^{th}$, 14$^{th}$ conduction states at the Γ point to these localized diagonal orbitals are calculated in the following rows. The Janak analysis and the LSDA+$U$ results on the change in the band energy due to +$U$ are listed in the last two columns.

Also note that the local spin density is completely determined from the occupied Bloch states, while the overlap between unoccupied Bloch states and occupied diagonal local basis ($n_{l\sigma} > 0.5$) is not necessarily zero. Indeed, our numerical results agree with our intuition that the $e_g$ states will be more delocalized and have smaller overlap with localized diagonal orbitals. Also, the empty major-spin $e_g$ states have small but noticeable overlap between the occupied major-spin diagonal orbitals, giving rise to the negative shift of CBM energy with increasing $U$, as confirmed from both our Janak analysis and DFT calculations (last two columns in Supplementary Table 4). Moreover, the $p$-dominant VBM state ($v_1$) at the Γ point has little overlap with those $d$-like localized diagonal orbitals, leading to negligible energy shift upon +$U$. For these reasons, the bandgap slightly decreases with increasing $U$.

**Supplementary References**